\documentclass[preprint]{aastex}

\begin{document}

\title{The Globular Cluster Systems of the Sculptor Group}

\author{Knut A.G. Olsen\altaffilmark{1}\affil{National Optical Astronomy Observatory,
CTIO, Casilla 603, La Serena, Chile\\ kolsen@noao.edu}}

\author{Bryan W. Miller\altaffilmark{1}} \affil{Gemini Observatory, Casilla 603, La
Serena, Chile \\ bmiller@gemini.edu}

\author{Nicholas B. Suntzeff\altaffilmark{1}\affil{National Optical Astronomy Observatory,
CTIO, Casilla 603, La Serena, Chile\\ nsuntzeff@noao.edu}}

\author{Robert A. Schommer\altaffilmark{1}\altaffilmark{2}\affil{National Optical Astronomy
Observatory, CTIO, Casilla 603, La Serena, Chile\\
rschommer@noao.edu}}

\author{John Bright\altaffilmark{3}\affil{Center for Astrophysics\\
jbright@head-cfa.harvard.edu}}

\altaffiltext{1}{Visiting Astronomer, Cerro Tololo Interamerican
Observatory, National Optical Astronomy Observatory, which is
operated by Associated Universities for Research in Astronomy, Inc.,
under cooperative agreement with the National Science Foundation}
\altaffiltext{2}{Deceased 2001 December 12.}
\altaffiltext{3}{Participant in the 2000 CTIO Research Experiences for Undergraduates program, which is supported by the National Science Foundation.}

\begin{abstract}
We use CTIO 4-m Mosaic II images taken with the Washington $CM$ and
Harris $R$ filters to identify candidate globular clusters in the six
major galaxies of the Sculptor group: NGC 45, NGC 55, NGC 247, NGC
254, NGC 300, and NGC 7793.  From follow-up spectroscopy with
Hydra-CTIO, we find 19 new globular clusters in NGC 55, NGC 247, NGC
253, and NGC 300, bringing the total number of known Sculptor group
globular clusters to 36.  The newly discovered clusters have
spectroscopic ages consistent with those of old Milky Way globular
clusters, and the majority are metal-poor.  Their luminosity function
closely resembles that of the Milky Way's globular clusters; their
metallicity distribution is somewhat more metal-rich, but this may be
the result of our color selection of candidates.  The mean
[$\alpha$/Fe] ratio in the clusters is $-0.2\pm0.3$, which is lower
than the Milky Way average.  The specific frequencies $S_N$ are
similar to those of other late-type galaxies.  However, if we
calculate the specific frequency using the $K$-band total magnitudes
of the host galaxies, we find values that are more than a factor of
two higher.  The kinematics of the globular cluster systems are
consistent with rotation with the \ion{H}{1} disk in each of the four
galaxies; however, only in NGC 253 is this result based on more than 
seven objects.  We suggest that the Sculptor group galaxies add
to evidence indicating that many of the first generation globular
clusters formed in disks, not halos.
\end{abstract}

\section{Introduction}
Observations of extragalactic globular cluster (GC) systems suggest a
variety of processes by which they may form.  For example, giant
elliptical galaxies may build up their GC systems in galaxy mergers
through the combination of the progenitor systems (e.g. Forbes et
al. 2000) and through the formation of new GCs in the merger
(e.g. Schweizer et al. 1996).  Galaxies such as the Milky Way and M31
likely formed many of their GCs from fragments in the halo (Searle \& Zinn 1978).  Parts
of GC systems may also have been built through the later accretion of
the GCs of dwarf galaxies (C\^{o}t\'{e} et al. 1998); indeed, the Sagittarius dwarf is falling into the Galaxy and being torn apart by tidal forces (Ibata et al.\ 1994), in the process contributing $\sim$4 globular clusters to the halo (Da Costa \& Armandroff 1995).   Each of these
processes may perhaps be explained through the common framework of
hierarchical structure formation (e.g. Beasley et al. 2002, Santos 2003, Kravtsov \& Gnedin 2003).

The prominent roles of merging and accretion in the GC formation
models suggest that it is very important that we measure the
properties of GCs in galaxies with lower masses, such as dwarf
galaxies and late-type spirals, in a range of environments.  In the
Local Group, the Large Magellanic Cloud (LMC) and M33 contain the
largest GC systems outside the Milky Way and M31.  The LMC's 13 old
GCs appear to be as old as the Milky Way's oldest clusters (to within
1 Gyr; Mighell et al.\ 1996, Olsen et al.\ 1998, Johnson et al.\
1999), and are metal-poor \citep{o91,suntzeff92}.  On the basis of
Monte Carlo simulations of their kinematics, \citep{s92} argued that
the LMC's oldest clusters follow the HI disk \citep{s92}.  If correct,
this result implies that the disks of some galaxies were already in
place by the time of GC formation, as predicted by cosmological
simulations (e.g. Kravtsov \& Gnedin 2003).  However, van den Bergh
(2004) argues that the LMC old cluster sample is insufficient to
rule out halo dynamics.
In M33, the 48 known old GCs reside
both in a disk and in a halo \citep{c02,s91}, while the bulk
have red horizontal branches, suggesting ages younger than those of Milky Way GCs \citep{s98}.  While these results are intriguing, global conclusions
about GC systems in low-mass galaxies clearly require larger
observational samples.

The Sculptor group is the Local Group's nearest neighbor; its brightest 
seven members have luminosities similar to that of the LMC
and M33 \citep {c97}.  The
small distances of the galaxies ($\sim$2 Mpc) make it possible
to study their GC systems with integrated photometry and
spectroscopy from the ground, and with resolved-star photometry from space.  Finding the GCs is challenging, however.  The expected
surface density of GCs is lower than that of background galaxies
and foreground stars, and the clusters are distributed over wide
fields.  Discovering the GCs thus requires a combination of wide-field imaging for candidate identification and follow-up spectroscopy for confirmation.
Searches in NGC 55 and NGC 253, using
photographic plates for candidate detection, have turned up a
number of GCs \citep{dc82,l83a,l83b,b86,b00,kim02}.  While these pioneering searches have confirmed the existence of Sculptor group GCs, their conclusions remain limited by their small samples.

In this paper, we report on the discovery of GCs
identified through a new wide-field CCD imaging (using the Washington filter set) and spectroscopic survey of six galaxies in the
Sculptor group, the details of which are described in \S2 and \S4.  We employ techniques similar to those described by \citet{b00} to identify candidate GCs; these are described in \S3.  In \S5, we describe our analysis of the spectral ages and metallicities, the photometric metallicities, the kinematics of the GC systems, and the Sculptor group GC luminosity function.  Our data allow us to perform the first detailed comparison between the GC systems of the Sculptor group and those of the Milky Way.

\section{Imaging observations and reductions}
We surveyed the six Sculptor group galaxies NGC 45, NGC 55, NGC 247,
NGC 253, NGC 300, and NGC 7793 with the CTIO 4-m telescope, Mosaic II
camera \citep{muller98}, and Washington $CM$ and Harris $R$ filters during the
nights 11-14 Nov 1999 (Table 1).  As its name implies, the Mosaic II
camera tiles the focal plane with 8 SITe 2K$\times$4K CCD chips, each
of which we read out through a single amplifier per chip with total
readout time of 160 s.  The spaces between the chips leave blank
columns and rows, each $\sim$15\arcsec ~wide.  In order to fill in
these gaps and to aid in cosmic ray removal, we typically used a
3-point dither sequence with 38\arcsec ~telescope offsets between
exposures.  
The images
were taken under $\lesssim1^{\prime\prime}$ seeing conditions; the last two nights were photometric, while the first was affected by some thin cirrus.  We set our exposure times so as to achieve
S/N$\sim$100 in $R$ and $S/N\sim50$ in $C$ and $M$ for GCs with
$M_R\sim-5.5$ at a distance of 2 Mpc.  The exposure times listed in
Table 1 were calculated based on the distances listed in \citet{c97}.
The distances to many Sculptor group galaxies have been revised since
the time of our observations \citep{karachentsev03}; we use these updated distances
throughout the analysis in this paper.

The Mosaic II data were reduced in IRAF through the MSCRED package
(Valdes 1997).  The reduction steps included removal of the effects of
electronic crosstalk between neighboring chips, subtraction of the
line-by-line median of the overscan regions of each chip, trimming to
remove extraneous pixels, subtraction of an averaged nightly bias
image, and division by averaged flat field frames produced each
afternoon by illuminating the white spot in the 4-m dome.  For the $C$
and $M$ filters, we found it necessary to correct for the differences
between these dome flats and flats produced from observations of the
dark sky.  Because our deep images all contain galaxies measuring
$\sim$10--20\arcmin ~across, we used Mosaic II $CM$ images of Galactic halo
fields taken by Mario Mateo and Robbie Dohm-Palmer during the nights
29 - 30 Oct 1999, and kindly provided to us by them, for the dark sky
illumination corrections.  These data were first processed and
flat-fielded using dome flats taken during the same run.  We then used
DoPHOT \citep{schechter93} to mask out the astronomical sources and combined the
masked images through a pixel-by-pixel median operation.  For the $M$
data, we median filtered the combined dark sky image with an 11
pixel-wide box before dividing it into our Sculptor group and standard
star images.  For the $C$ data, we found that the differences in
flat-field features between the dark sky and dome-illuminated flats
were too great to be adequately corrected with a median-filtered dark
sky illumination frame.  We thus divided the $C$ dark sky illumination
frame into our data without any filtering; the $C$ images are thus
effectively only flattened with a dark sky flat frame.  The total
counts in the $C$ dark sky flat were $\sim$5000 ADU, or $\sim$10000
electrons; the flat field accuracy for the $C$ filter should thus be
$\sim$1\%.

We used an observation of a field near the galaxy NGC 55 to refine the
astrometric solution that is automatically written to Mosaic II image
headers.  The astrometric solution, which was derived by matching
sources from the USNO-A2 catalog \citep{monet98} to sources found in our image
through the task MSCTPEAK, consisted of a fourth order polynomial in
two dimensions.  After applying the solution to all of our data, we
found typical RMS deviations of 0\farcs3 when comparing the positions
of sources in our images with those from the USNO-A2 catalog.

For the purpose of selecting candidate globular clusters, we produced
single combined images for each Sculptor group galaxy through each
filter.  We first subtracted a single value of the sky background from
each image, which we determined from the mode of the pixel intensity
distribution taken by excluding pixels contaminated by light from the
Sculptor group galaxies.  We next created single-extension images from
each of the multi-extension processed images using the task MSCIMAGE,
which stitches together the images from individual chips using the
astrometric information contained in the image headers and resamples
the images to a uniform pixel scale.  We then aligned the images of
each galaxy and filter with the tasks MSCZERO and MSCCMATCH, scaled
the individual image intensities with MSCIMATCH, and combined the
images with MSCSTACK.  Figure 1 shows a montage of the Sculptor group
galaxy images.

For the purpose of measuring accurate photometry of objects found in
the stacked images, we also worked with the individual reduced images on a
chip-by-chip basis.  In preparation for the single-CCD photometry, we accounted
for the variation in effective pixel area produced by geometric
distortion in the camera by multiplying by a pixel area image.  This
image was produced with the MSCRED task MSCPIXAREA; the task uses the
astrometric information in the image headers to calculate the
transformation of pixel areas to areas observed on the sky.  Because
MSCIMAGE resamples the images to a uniform pixel scale, this step was
unnecessary for the production of the stacked images.  Finally, we split
each of the image mosaics into eight images, one for each CCD
detector.

\section{Photometry and selection of candidate globular clusters}
The 36$^\prime\times36^\prime$ field
of our Mosaic II images is likely to capture the majority of Sculptor group GCs in a single snapshot of each galaxy.  
We selected candidate globular clusters from the deep, stacked images,
using criteria based on photometry and object morphology.  We
conducted photometry on the combined galaxy images using the program
SExtractor \citep{bertin96}.  For object detection, we used a 3$\times$3-pixel
Gaussian filter with FWHM of 1.5 pixels.  For all objects containing
at least five pixels 5$\sigma$ above the local background, we measured
the light from all contiguous pixels with values 5$\sigma$ above the
background.  Because the candidate GCs have a range of profiles, we
used these ``isophotal'' magnitudes in place of fixed apertures.
Profile-fitting photometry, such as done by Seth et al.\ (2004), would be necessary for measuring GC candidates
projected against the crowded galaxy disks.  However, in order to
limit the number of candidates for spectroscopic followup, to mitigate
against the effects of internal absorption by dust, and to simplify
the photometric analysis, we chose to exclude the most heavily crowded
regions from our analysis.  The regions for exclusion were selected by
eye (see Figure 1), and correspond roughly to the region where
SExtractor failed to distinguish groups of point sources as composed of separate objects.

We calibrated the Sculptor group photometry through nightly observations of
the \citet{landolt92} fields that were standardized to the Washington
system by \citet{geisler96}.  The last two out of the three nights were photometric, while the first was useful but non-photometric.  We typically observed a standard star field
every $\sim$2 hours, with the standards usually centered on CCD \#6.
We observed standards over the magnitude range $10.6<m_{T1}<14.8$, the color range $0.64<(C-T1)<2.64$, and the range of airmass $1.15<X<2.0$. 
Over the course of our three nights, we also observed the SA 98 field
with standards placed on each of the 8 detectors in turn, so as to
allow measurement of the individual zero points and color terms of
each detector.  We performed aperture photometry on the standard star
frames using DAOPHOT \citep{stetson87} and a sequence of apertures, the smallest having a diameter of 6\arcsec and the largest a diameter of 15\farcs75, close to the size used by Landolt (1992).  We used DAOGROW to compute growth curves from the multi-aperture photometry, to select the best aperture for each star, and to extrapolate the photometry to a 30\arcsec  aperture.  For the 15\farcs75 aperture, the size of the extrapolation was $\sim-0.001$ magnitudes, while for the 6\arcsec aperture it was $\sim-0.02$ magnitudes.  Given the total instumental magnitudes for the standard stars, we solved for the coefficients of the photometric
transformation equations of the form (e.g. \citet{stetson88}):
\begin{eqnarray}
c = C + A_0 + A_1(C-M) + A_2X \\
m = M + B_0 + B_1(C-M) + B_2X \\
r = T_1 + C_0 + C_1(C-T_1) + C_2X
\end{eqnarray}
where $X$ is the airmass.  As indicated by the third equation, we transformed our $R$ filter instrumental magnitudes to Washington $T_1$ standard magnitudes; as discussed by Geisler (1996), the $R$ filter is a close match to the $T_1$ filter, and has higher throughput.  Table 2 gives the transformation coefficients for each detector.

In order to tie the SExtractor photometry to the standard system, we conducted aperture photometry of bright, uncrowded Sculptor group sources using DAOPHOT and the individual images.   This photometry was perfomed in a fashion similar to that used for the standard star photometry, and is thus easily tied to the standard system.  We used a single aperture with diameter 8\farcs1 arcsec; the standard star photometry growth curves indicate that this aperture contains $\sim$99\% of the total light from point sources.  The measured magnitudes were transformed to $C$, $M$, and $T_1$ using equations 1--3, adopting the zero points and color terms calculated for each individual chip.  All of the standard magnitude measurements were then averaged to produce a final magnitude for each bright Sculptor group source.  We then matched these sources to objects found in the analysis of the stacked images, and applied an average zero point to the SExtractor $C$, $M$,and $T_1$ photometry of each galaxy.  Because the variations in the chip-to-chip zero points and color terms are small, we expect that the systematic error in the photometric calibration is no larger than $\sim$3--5\%.

\subsection{Selection of candidate globular clusters}
We corrected the $CMT_1$ magnitudes for reddening using
the Galactic foreground measurements of \citet{sfd98}, with $E(B-V)$
transformed to the Washington system following \citet{harris79}; the typical
$E(B-V)$ towards the Sculptor group is $\sim$0.02 magnitudes.  Because our candidate globular clusters all lie far outside the host
galaxy disks, we expect that the effects of internal dust extinction
are negligible.  Using the \citet{karachentsev03} distances to compute $M_{T_1}$ from our photometry, we used the colors and magnitudes spanned by Milky Way globular clusters to help select candidates for spectroscopic followup; Milky Way globular clusters have
$1.0\lesssim(C-T_1)_\circ\lesssim1.9$ \citep{geisler90} and $-10.7 \lesssim M_{T_1} \lesssim -6$, with a peak in the GC luminosity function at $M_{T_1}\sim -8$ \citep{h96}.  Figure 2 shows the $C-T_1,T_1$ color-magnitude diagrams of our fields, with the GC selection limits indicated.  In NGC 55, NGC 247, and NGC 300, an unfortunate error in a preliminary photometric calibration resulted in the color selection limits being $\sim$0.2 magnitude redder than is desirable; nevertheless, our candidate lists are representative of the colors and magnitudes of the bulk of Milky Way GCs.

Next, we used the shapes and areas of the isophotes computed by
SExtractor to make a further selection of likely globular cluster
candidates.  We guided our selection criteria based on 1) the
properties, as measured from our images, of 10 of the 14 NGC 253
globular clusters spectroscopically identified by \citet{b00}, and 
2) the simulated appearance of Milky Way globular clusters
in our images.  From the known NGC 253 globular clusters, we
established the rather loose criterion that globular cluster
candidates should have isophotal ellipticities $e<0.4$, where $e=0$
represents a circular isophote.  To simulate the appearance of Milky
Way globular clusters at the distances of the Sculptor group galaxies,
we used DAOPHOT to measure the point-spread function (PSF) from stars
found in each of our $R$ images.  We then used the structural
parameters from \citet{h96} to produce model \citet{king66} profiles for
the Milky Way clusters, which we convolved with our PSFs.  From the
convolved profiles, we calculated the area contained within the
isophote lying 5$\sigma$ above the background, as measured from our
images.  Figure 3, which is similar to Fig.\ 6 of \citet{b00}, shows the isophotal areas and $R$-band fluxes of objects found
in our Sculptor group images compared with our simulated results for
Milky Way clusters.  In this diagram, we expect point sources to follow a tight, nearly linear sequence in $\log$(area) vs.\ $\log$(flux); stars with larger flux have larger area, because more of the wings of the PSF rise above the surface brightness limit.  We expect background galaxies to follow a looser sequence, with larger area for a given flux than point sources.  Globular clusters in the Sculptor group generally fall in between the point source and galaxy sequences, although clusters in the more nearby Sculptor group galaxies, such as NGC 55, have significant overlap with distant background galaxies in the diagram.
While the most compact Milky Way globular
clusters are indistinguishable from stars, many of the globular
clusters would be visibly resolved.  We considered all objects that
obeyed the aforementioned color, magnitude, and ellipticity limits and
that passed visual examination (to eliminate image artifacts and
obvious background galaxies) as viable spectroscopic targets.
We then used the detection of broadening to trim the candidate list to $\sim$100 objects per galaxy, which is a manageable number for followup with multi-object spectrographs.  In the case of NGC 55, there were few enough candidates that we were able to include all point sources and extended objects in the approximate magnitude and color range of globular clusters in the candidate list.  Table 3 lists the properties of our candidate globular clusters.

\section{Spectroscopic observations and reductions}
We observed GC candidates in NGC 55, NGC 247, NGC 253, and NGC 300 with the Hydra-CTIO multi-fiber
spectrograph \citep{barden98} during the nights 31 Oct - 3 Nov 2000 and 17 - 20
Oct 2001.  The nights in 2000 were almost completely clouded
out, and provided no useful data on the globular cluster candidates; they did, however, provide useful data on a number of bright Lick index \citep{worthey94} standard stars.  The nights in 2001 were all
clear.  Table 1 summarizes our
spectroscopic observations.

Hydra-CTIO has 138 2.0 arcsec-diameter fibers that may be
robotically positioned with high accuracy on a magnetic plate spanning
a 40\arcmin-diameter circular field, providing an excellent match to
the field size of Mosaic II.  For the nights in 2001, we typically positioned $\sim$80 object
fibers, $\sim$20 sky fibers, and $\sim$3 guide fibers per new field
within 20 minutes. We used the atmospheric dispersion corrector, which
compensates for the effects of differential atmospheric refraction.
The fiber positioning was done on the flat surface of the magnetic
plate; however, because the focal plane is curved, all observations
were taken with the plate made curved by a vacuum chamber attached to the
plate.  Hydra-CTIO's fibers feed a bench spectrograph located in an
isolated room beneath the dome floor, a setup which allows for the
stable measurement of radial velocities.  The spectra taken in 2001 were imaged by a
Schmidt camera with 400 mm focal length onto a SITe 2K$\times$4K CCD,
which we used in its low-gain (0.84 e$^-$/ADU), low read noise ($\sim$3 e$^-$)
setting.  The resolution of the camera is such that unless a slit
plate is used to increase spectral resolution, the spectral profile is
oversampled by a factor $>$2; we thus chose to bin the CCD by two
pixels in the spectral direction, which reduced the effects of read
noise and shorthened readout time, while preserving spectral
resolution.  We used the KPGL1 grating over the wavelength range 3700
-- 5900\AA, which provided $\sim$2\AA ~spectral resolution.  The spectra of bright Lick standard stars, taken in 2000, were imaged with a 229mm Schmidt camera and Loral 3K detector.  While this camera had poorer image quality and read noise than the 400mm camera used in 2001, the grating, wavelength setting, and spectral resolution were all the same.

Calibration data taken at the telescope included bias frames, dark
frames to monitor possible light leaks entering the open bench
spectrograph, twilight sky frames, flat field frames taken while
pointed at the illuminated dome white spot, flat field frames taken
during the night while illuminating a projector screen located inside
the telescope chimney, and comparison lamp exposures (using Hydra-CTIO's He/Ne/Ar emission line sources) taken either
immediately before or after an observation.  We also took some
exposures during the night with all fibers pointing at blank sky, to
allow us to experiment with the sky subtraction procedure.  During the
second afternoon, we opened the dome and placed a diffusing filter in
front of the row of fibers in the spectrograph; exposures taken in
this configuration were relatively uniformly illuminated with light,
and were used to construct a two-dimensional flat field image (``milk flat'').

Our observing routine in 2001 generally started in the afternoon with a
sequence of 11 bias frames and a 10-minute dark exposure, after which we placed all of the available
fibers in an $\sim$27\arcmin ~diameter circle centered on the field.  After
pointing the telescope at the white spot in the dome and turning on
the quartz lamps, we then took a sequence of three 10-minute
flat field exposures, preceded and followed by a comparison lamp exposure.  Next,
we configured the fibers for the first Sculptor group field for that
night, with the exception of the last night, when we started
observations with a field in the Milky Way globular cluster 47 Tuc.
Our evening twilight sky exposures were taken in these configurations,
followed by projector flat and comparison lamp exposures; on the last night, we took morning twilight sky exposures in the NGC 247 configuration.  We took
observations of K giant radial velocity standards and
spectrophotometric standards at the beginning and end of each night
(again followed by projector flat and comparison lamp exposures), for
which we placed a single fiber in the center of the field.  The
remainders of the nights we exposed on a single Sculptor group field
per night, with occasional tweaking of the fiber positions to account
for changes in airmass and frequent interruptions for project flat,
comparison lamp, and blank sky exposures.  

The data were processed in IRAF, using CCDPROC to subtract a
polynomial fit to the overscan region, trim extraneous pixels from the
image edges, and subtract the averaged nightly bias frame.  Division
of selected images by the processed milk flat image was found not to
reduce the spectral noise, and so was not performed in the final
reduction.  Our long exposures contained a large number of cosmic
rays; we used the routine L.A.Cosmic (van Dokkum 2001) to remove these
from the individual object images with good results.

The spectra were extracted with the IRAF script DOHYDRA.  We used the
projector flats taken during the night to trace the individual fiber
profiles and to set the extraction apertures, the afternoon dome flats
to provide flat field vectors for each fiber, the twilight sky
exposures to assign the fiber-to-fiber throughput values, and the
nearest available He/Ne/Ar comparison lamp exposures to set the
wavelength scales for each fiber.  The trickiest aspect of the
extraction procedure was the sky subtraction.  We carried through the
extractions using two sky subtraction options: 1) using the $\sim$20
sky fiber spectra per field, scaled to the throughput value for each
fiber, we subtracted an average sky spectrum from each object
spectrum, and 2) using the nearest dark sky observation, we subtracted
the sky image from the on-object image before extracting the
one-dimensional spectra.  In practice, we found that the first option
subtracted the sky as well as the second, while the second option
slightly increased the random noise in the one-dimensional spectra; we
thus chose to perform the sky subtraction using the sky fibers rather
than the dark sky exposures.  Following the extraction, we
interpolated the spectra to a common linear wavelength scale.  Because the
continuum shapes varied from exposure to exposure with changing
atmospheric conditions, we also matched all of the spectra of a given
object to a single continuum before averaging them into single spectra
for each object.

\subsection{Radial velocities}
We measured velocities for each Sculptor group object by
cross-correlation \citep{tonry79} with our spectra of stars with
known radial velocities.  The cross-correlation was done with the IRAF
task FXCOR.  First, we used FXCOR to perform continuum subtraction of the
object spectra.  Next, we applied a ramp filter to dampen the highest
and lowest frequency Fourier components; if left unfiltered, these frequencies
produced broad features that masked the narrow cross-correlation
peaks.  We fit the narrow cross-correlation peaks with Gaussians, which
produced typical internal velocity errors per measurement of 30 km s$^{-1}$ for objects with
spectra similar to the templates.  We cross-correlated each of the object spectra with 7 template spectra (taken over the course of our four nights) of objects with known velocities.  When the measurements using each of the 7 templates were averaged, we found typical standard deviations of $\sim$7 km s$^{-1}$, as expected by $\sqrt{N}$ statistics.
For a number of objects, no
significant cross-correlation peak was found.  Most of these objects
contained emission lines at wavelengths other than the rest
wavelengths of common strong lines (indicating that they are background galaxies), some had spectra that were too
weak for analysis, while one had the spectrum of an HII region at the
velocity of NGC 55.  Table 3 summarizes the radial velocity results.

In order to assess our true velocity accuracy, we cross-correlated the
spectra of the K4III star HD 223311 that were observed on different
nights.  From the 12 object/template combinations, we found a standard
deviation of 5 km s$^{-1}$.  Thus, we think that a velocity error of
$\sim5-10$ km s$^{-1}$ is an honest assessment of our true errors.

\section{Results}
\subsection{Identifying the Sculptor group star clusters}
The measured velocities easily remove contaminating background
galaxies from our sample of globular cluster candidates.  As a first step to
identify likely foreground stars from objects in the Sculptor group,
we used a combination of our measured velocities and the measured FWHM of our objects compared to the median FWHM of all point-like sources.
Figure 4, which plots velocity vs. the ratio of the FWHM of
the object to that of typical point sources in the images, demonstrates our
technique.  Because the Sculptor group lies near the South Galactic Pole, foreground stars have an average heliocentric velocity
near 0 km s$^{-1}$ and, as calculated for 568 stars in the direction of
the Sculptor group taken from the General Catalog of mean radial velocities \citep{rv00}, a velocity dispersion of
40 km s$^{-1}$.  This is similar to the vertical velocity disperion of the thick disk, $\sigma_z\sim45$ km s$^{-1}$ (Gilmore, Wyse, \& Kuijken 1989).  The systemic velocities of NGC 55, NGC 247, NGC 253, and NGC 300,
on the other hand, are in the range 125 -- 245 km s$^{-1}$ \citep{c90,p90,p91a,p91b}.  

In addition,
stars have profiles that vary by only $\sim$10\% in their widths
compared to the median PSF, as measured from the distribution of
objects in Figure 4.  Figure 4 contains a number of somewhat broader
objects, many of which have velocities similar to the systemic
velocities of the Sculptor group galaxies.  We selected as
likely Sculptor group star clusters those objects that either had
velocities in excess of 2$\sigma$ of the mean foreground star
velocity, or that were 2$\sigma$ broader than the mean PSF of the
image in which they were found; the majority of our selected clusters have velocities or broadening factors in excess of 3$\sigma$ compared to stars.  We were of course unable to identify compact GCs that
have $v_{\rm hel}<$ 100 km s$^{-1}$, but we expect these to be relatively small
in number.  

We found that 31 objects obeyed our candidate selection criteria, and
considered these excellent Sculptor group star cluster candidates.
However, we needed to consider the possibility of further contamination
from Galactic foreground stars.  In our observed color and magnitude
range, there is possible contamination from dwarf stars at distances
of $\sim2.5-6$ kpc, and by giants at distances $\gtrsim$25 kpc.  Because
the Sculptor group galaxies lie near the South Galactic Pole, we
expect that contaminating giants would lie in the halo, while the
dwarf stars would be members of either the thick disk or halo.  Since at the magnitudes of interest halo giants are outnumbered by foreground dwarfs by $\sim$100:1
\citep{morrison00,majewski00}, dwarfs are the likeliest stellar contaminants.  We thus used the
gravity-sensitive features in our spectra, in particular the Mg$b$ and
MgH features at $\sim$5200\AA ~as well as Ca4227, to identify
foreground dwarfs.  Figure 5 shows the spectra of all of the
candidates.  Seven of the
spectra--those of NGC 55 candidates \#1, \#2, \#3, \#5, and \#6 and
NGC 300 candidates \#2 and \#5--have the strong Mg$b$ and MgH
absorption of dwarf stars.  These spectra can be
best fit by Kurucz (1993) models appropriate for solar-metallicity main
sequence dwarfs ($\log g\sim$5.0), rather than lower-gravity models
appropriate for integrated spectra of star clusters.  Moreover,
examining their position in the diagnostic diagram of Figure 4, we found
that 3/7 lie near or below the 3$\sigma$ velocity limit, while the
other four are broadened objects in NGC 55 that lie on the edge of the
Mosaic 2 image and have near-zero heliocentric velocity.  We thus removed these seven objects from the sample and
considered the remaining 24 objects as likely Sculptor group star clusters.  Table 4 summarizes the properties of the Sculptor group GCs.

\subsection{Measurement of Lick indices}
We measured Lick/IDS (Brodie \& Huchra 1990) indices, as redefined by
Trager et al.\ (1998), for our star clusters and for the Lick/IDS
standards HD 3567, HD 22879, HD 23249, HD 196755, HD 218527, HD
219449, and HD 222368.  We were primarily interested in the Balmer
indices, Mgb, Fe5270, Fe5335, and Ca4227; the Balmer indices are
useful as age indicators, while the combination \newline
[Mg/Fe]$^\prime = \sqrt{{\rm Mg}b(0.72\times{\rm Fe}5270 +
0.28\times{\rm Fe}5335)}$ (Thomas, Maraston, \& Bender 2003) provides
a tracer of total metallicity that is independent of $\alpha$/Fe.

Because our spectra are of much higher resolution than those used to
define the Lick/IDS system, we smoothed our spectra to the Lick/IDS
resolution before measuring the equivalent widths of the indices.  To
do this, we smoothed our spectra with a variable-width Gaussian having
FWHM linearly interpolated from Worthey \& Ottaviani (1997)'s Table 8,
which lists the Lick/IDS resolution as a function of wavelength.  We
then measured the equivalent widths following the prescription of
\citet{trager98}.  The index definitions were first shifted to
account for the radial velocity shifts measured in section \S4.1;
distortions in the wavelength scale were previously removed by our
spectral reduction procedure.  For each index, we then fit a straight
line between the fluxes of the blue and red pseudocontinuum regions.
The equivalent widths were measured by integrating the ratio of the
observed flux to that of the local continuum fit over the wavelength
regions of the indices.  We computed errors in the equivalent widths
by adding in quadrature the contributions from Poisson noise and the
error introduced by our radial velocity uncertainities.  Comparison of
our measured equivalent widths and the standard values for the
Lick/IDS standard stars provide an external measure of our errors.
Figure \ref{fig7} shows that the scatter in the residuals of the measured and
standard index values is much larger than our internal estimates.  We
thus adopt the standard deviation of the residuals, which are
$\sim$0.2\AA, as the minimum true errors in equivalent width.  Worthey
et al.\ (1994) note that flat-fielding errors generally provide the
dominant contribution to the errors in equivalent width; as this
source of error is difficult to estimate, we did not account for it in
our internal error budget.  Figure \ref{fig7} shows that of all the indices, only Ca4227 shows a bias in the equivalent width that is appreciably larger than the internal scatter.  We subtracted this $\sim$0.2\AA ~bias from the Ca4227 measurements.

Our observations of Lick standard stars were taken with much higher
signal-to-noise ratio (S/N$>100$) than were the Sculptor group
observations.  To evaluate the effect of lower S/N on the measurement
of the equivalent widths, we added various levels of Gaussian random
noise to the Lick standard spectra and remeasured the equivalent
widths.  We found that above
$S/N\gtrsim25$, the effect of lower S/N is to increase the scatter in
equivalent width by $\lesssim$10\%, which is typically equal to or
smaller than the 0.2\AA ~intrinsic error.  Below $S/N\lesssim25$, the
error introduced by lower S/N increases dramatically, approaching 50\%
at S/N$\sim$10.  We thus analyzed the Lick indices of only those six Sculptor
group star cluster spectra with S/N$\gtrsim25$ independently.  For the
remaining 18 lower S/N spectra, we shifted each to zero heliocentric
velocity and measured the Lick indices of the single combined spectrum
(Figure \ref{fig8}).  Table 5 summarizes our Lick index measurements.

\subsection{Ages and metallicities of Sculptor group star clusters}
As mentioned previously, the combined index [Mg/Fe]$^\prime$ (Thomas
et al.\ 2003) is a useful total metallicity indicator, while the
Balmer indices provide an indication of age.  Figure \ref{fig9} shows the
Balmer indices and [Mg/Fe]$^\prime$ for our 6 high-S/N spectra of
Sculptor group candidate star clusters and for our combined spectrum
of 18 clusters, compared to the single stellar population models of
Thomas et al.\ (2003).  

The comparison of the points with the models
indicates that all of the star clusters of the Sculptor group have
[Fe/H]$\lesssim-1.0$.  All of the candidates also have H$\beta$ strengths
consistent with old age ($\sim$12 Gyr), except perhaps one, NGC 247
\#1, which may be intermediate in age, or has a very blue
horizontal branch (cf. Beasley, Hoyle, \& Sharples 2002).  Using H$\delta_F$ or
H$\gamma_F$ (Worthey \& Ottaviani 1997) in place of H$\beta$ yields
similar results; most, if not all, clusters are compatible with old
ages.  As happens even in cases with excellent S/N spectra (e.g. Beasley, Hoyle, \& Sharples 2002), some of the points fall outside the ranges predicted by the models; one might thus argue that we know little about the absolute ages and metallicities.  However, given the small dispersion in [MgFe]$^\prime$ and, to a lesser extent, the Balmer indices, we conclude that the metallicity and age range spanned by the clusters is small.  Moreover, the qualitative agreement of our measured indices with those measured by Beasley, Hoyle, \& Sharples (2002) for the LMC's old globular clusters argues that our star clusters are likely old, metal-poor Sculptor group globular clusters.

We also used the Lick index measurements to investigate the
[$\alpha$/Fe] element abundance ratios of the clusters.  Figure \ref{fig9}
shows our measured Mg$b$/$<$Fe$>$ ratios compared to [MgFe]$^\prime$,
where $<$Fe$>=0.5\times($Fe5270 + Fe5335$)$, compared to the Thomas et
al. (2003) predictions.  The individually measured clusters are
consistent with $-0.3\pm0.15<[\alpha/{\rm Fe}]<0.0\pm0.15$, while the
[$\alpha$/Fe] ratio of the combined GC spectrum appears to be
$\sim$0.15$\pm0.1$.  The average and standard deviation of our
Mg$b$/$<$Fe$>$ measurements are 0.9 and 0.4 respectively, which
corresponds to [$\alpha$/Fe]$\sim-0.2\pm0.3$ in the Thomas et
al. (2003) model.  This value is low compared to that of Milky Way
globular clusters, which have [$\alpha$/Fe]$\sim$0.3 \citep{kraft94}.

How do the spectroscopic metallicities agree with those expected from
our Washington photometry?  We used the relationship of Harris \&
Harris (2002), which is based on the Harris (1996) catalog of Galactic
globular clusters and \citet{harris77}'s $CMT_1T_2$ photometry, to derive [Fe/H] from
our $(C-T_1)_0$ colors.  Figure \ref{fig10} shows the comparison between the
metallicities expected from $(C-T_1)_0$ and those from [MgFe]$^\prime$.
The agreement between [Fe/H]$_{\rm [MgFe]^\prime}$ and
[Fe/H]$_{(C-T_1)_0}$ is within the errors; we thus feel justified in
providing estimates of the abundances of each of the individual 24 globular
cluster candidates through our Washington photometry in Table 4.

\subsubsection{Comparison with the Milky Way}
Figure \ref{fig12} shows the luminosity function (LF) and [Fe/H]$_{C-T_1}$
distribution of the combined Sculptor group GCs compared to those of the Milky Way \citep{h96}.  The Sculptor group distributions include the two NGC 55 GCs discovered by \citet{dc82} and 6 out of 10 of the \citet{b00} GCs for which we have photometry but no spectra; the photometric measurements are listed at the end of Table 4.  Three of the \citet{b00} GCs were found to have $C-T_1$ colors either too blue or too red compared to Milky Way GCs, and are not included in Figure \ref{fig12}.  The LFs appear similar in the Milky Way and in the Sculptor group.  Applying a Kolmogorov-Smirnov (K-S) test to the
unbinned luminosity distributions, we found that the LFs have a $\sim$30\% probability of being drawn from the same parent distribution.  While the Sculptor group [Fe/H] distribution appears somewhat more metal rich, this difference may be the result of our inadvertent exclusion of GC candidates with $1.0 \lesssim (C-T_1)_\circ \lesssim 1.2$.  We thus conclude that there is no strong evidence to suggest different LFs or [Fe/H] distributions in the Sculptor group and Milky Way.

\subsection{Kinematics of the globular cluster systems}
To analyze the kinematics of the Sculptor group globular cluster systems, we combined our sample of GCs with those found by \citet{b00} in NGC 253 and NGC 55 and by \citet{dc82} in NGC 55.  
Figure \ref{fig11} shows the velocities of the GCs compared to the HI rotation curves \citet{c90,p90,p91a,p91b}, where we have projected their rotation curves to the line-of-sight using their derived inclinations.  To test for possible rotation in the GC systems, we fit the error-weighted GC velocities with a sinusoid of the form:

\begin{equation}
v_{\rm los} = v_{\rm sys} + v_{\rm rot,proj} \times \sin(\theta - \theta_{\rm rot})
\end{equation}

We computed two sets of solutions, one in which we allowed $v_{\rm
sys}$ be a free parameter and one in which we fixed $v_{\rm sys}$ to
the value derived from the \ion{H}{1} observations; the solutions are
summarized in Table 6.  For NGC 253, these two fits were nearly
identical, while we concluded that the other galaxies do not have
enough GCs to warrant the extra degree of freedom introduced by allowing $v_{\rm sys}$ to vary, as witnessed by the
large discrepancies in the systemic velocities calculated from the GCs vs. those from the \ion{H}{1}.  In all of the galaxies,
we found that the position angles of the fits to the GC velocities
agree closely with the position angle of the HI rotation curve.
Indeed, Monte Carlo simulations show that there is only a $\sim$5\%
probability that such closely aligned position angles could occur by
chance in NGC 55, NGC 247, and NGC 253, with $\sim$15\% probability in
NGC 300.  We thus suggest that the globular cluster systems of the
Sculptor group appear to have significant rotation, with $v/\sigma\sim
1-2$.  However, for NGC 55, NGC 247, and NGC 300, which have only a
few identified GCs, this conclusion is only tentative.  We
also acknowledge that we likely have missed those compact GCs that have
approaching velocities $\lesssim$100 km s$^{-1}$, which may also
affect these conclusions.

In NGC 55, the projected radii of the GCs all lie inside the outer
boundary of the \ion{H}{1}, while in NGC 300, the most distant cluster
lies $\gtrsim$2 kpc beyond the limit of the \ion{H}{1}.  Taking the
rotation solutions at face value, we find that the NGC 55 and and NGC 300
GC systems are consistent with having circular orbits and flat rotation curves out to radii of 7.8
and 12.7 kpc, respectively, in agreement with the \ion{H}{1}
kinematics.  In NGC 253, the GCs lie at projected radii of up to 22
kpc, a factor of 1.6 larger than the outer extent of the \ion{H}{1},
once the distance used by \citet{p91a} is adjusted to agree with
\citet{karachentsev03}.  Assuming that the NGC 253 GCs all lie in the
plane of the disk, we find deprojected radii of up to 65 kpc.  As seen in Fig. \ref{ngc253vel}, the GC system rotational velocity is a factor of $\sim$2 lower than that of the \ion{H}{1}; indeed, the sum of
the rotational kinetic energy and energy in random motions in NGC 253
falls a factor of 5$\pm$2 short of that expected from a flat rotation
curve extending to a radius of $>$15 kpc.  A possible explanation
may be that we have missed a number of compact GCs with low
line-of-sight velocities.  However, the close agreement between the
systemic velocities derived from our GCs and from the HI implies that
we must then have missed an equally large number of GCs with high
recessional velocities, for which we have no explanation.
  The only other explanation is that the NGC 253 GC sample suffers from asymmetric drift, in which the clusters we observe (which were selected to lie outside the disk of NGC 253) are on elliptical disk orbits with guiding centers preferentially located inside their observed radii.  In an exponential disk, van der Marel et al.\ (2002) find that $v_{\rm circ}^2 = v^2 + R/R_d\sigma_v^2$, where $R_d$ is the disk scale length; we can thus use the NGC 253 GC kinematics to estimate the disk scale length of the GC system.  Taking $v_{\rm circ}=224$ km s$^{-1}$, $<v>_{\rm GC}=78\pm18$ km s$^{-1}$, $<R>_{\rm GC}=27\pm18$ kpc, and $\sigma_v=76$ km s$^{-1}$, we find $R_{d,GC}=3.5\pm2.4$ kpc.  For comparison, \citep{p91b} measured the \ion{H}{1} disk scale length to be 2.4 kpc; thus, the scale length of the NGC 253 disk appears to have remained unchanged, within a factor of $\sim$2, between the time of GC formation and the present.

\subsection{Specific frequencies}
\begin{figure}
\epsscale{0.8}
\plotone{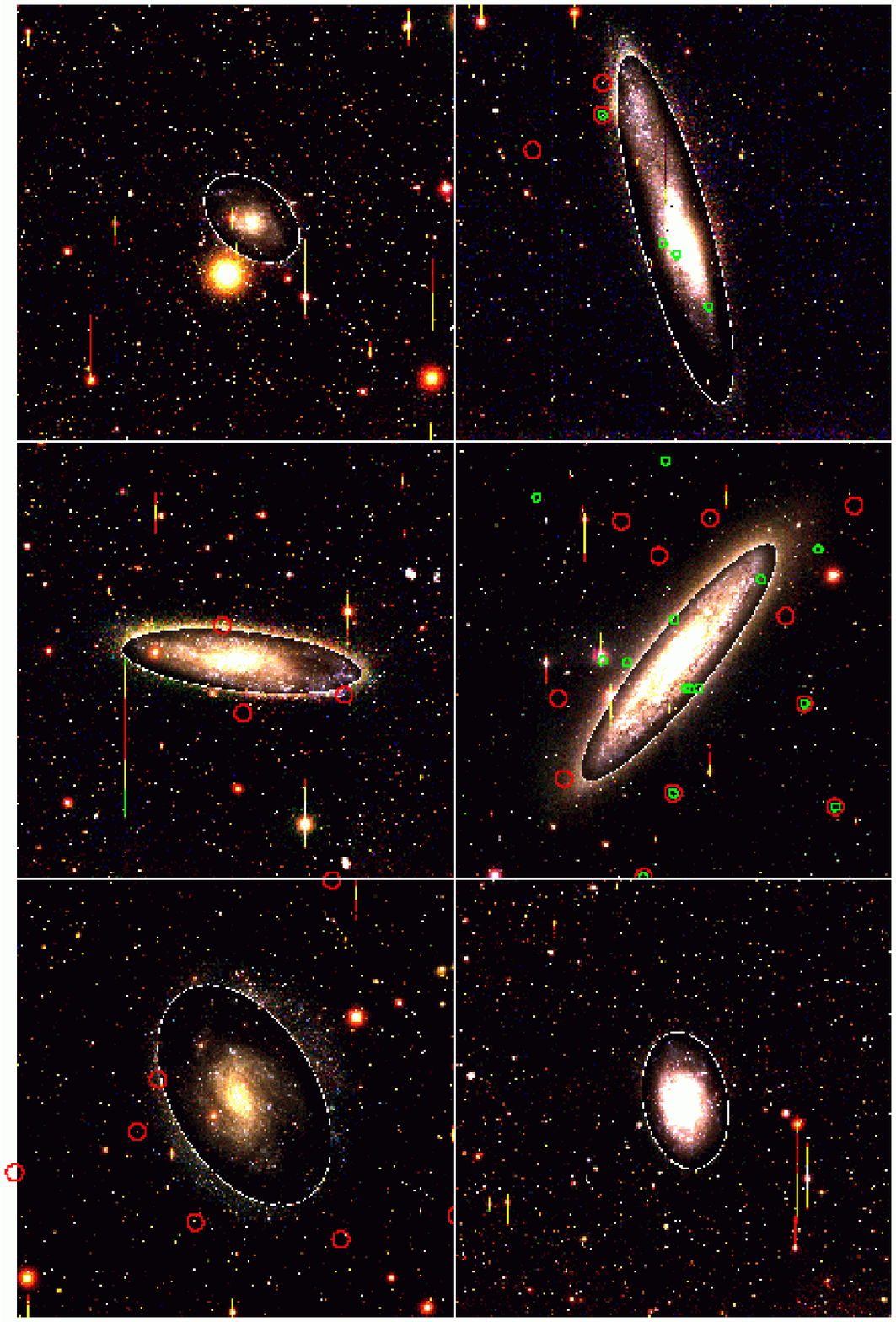}
\caption{The Sculptor group galaxies NGC 45, NGC 55, NGC 247, NGC 253,
NGC 300, and NGC 7793 (from top left to lower right) as seen with
the CTIO 4-m Mosaic II camera.  The red, green, and blue color planes
contain the combined images taken through the $R$, $M$, and $C$
filters, respectively.  The white ellipses mark the regions outside of
which we selected globular cluster candidates; inside the ellipses,
the images have been divided by a factor of 20 to avoid saturation of
the colors.  The red circles mark objects identified as globular
cluster candidates in our survey that we have confirmed through spectroscopy; the green circles mark the star clusters found by \citet{dc82} and \citet{b00}.}
\label{fig1}
\end{figure}

Comparing the properties of the Sculptor group GC populations with those of other galaxies rests on our ability to estimate the completeness of our GC samples.  Some GCs may have been lost because they lie either outside our images or within the elliptical masked regions of Figure \ref{fig1}, while others could fall outside our color and magnitude selection region.  We also did not obtain useful spectra of all of the objects identified as GC candidates, such that some GCs may remain undiscovered within our survey area.

We calculated the completeness of our Sculptor group GC samples using the Milky Way as representative of the complete GC population.  First, we placed the Milky Way GCs in the \citet{h96} catalog at the distances of the Sculptor group galaxies and calculated the fraction falling outside our images or within our masked-out regions.  Second, we multiplied this fraction by the fraction of Milky Way GCs from \citet{harris77} that obeyed our color and magnitude selection criteria.  Finally, we multiplied by the fraction of GCs for which we obtained useful spectra, assuming that the GCs for which we do not have spectra have the same luminosity function as those for which we do have spectra.  Our estimated completeness fractions are listed in Table 7.

The traditional measure of a galaxy's globular cluster content, the
specific frequency $S_N = N_{\rm GC}\times10^{0.4(M_V+15)}$, where $N_{\rm GC}$ is the total number of GCs and $M_V$ is the $V$ absolute magnitude of the host galaxy, is not
well-suited to late-type galaxies because of the presence of
populations much younger than those that formed the globular clusters that contribute to the total galaxy luminosities
(e.g. Harris 2003).  Nevertheless, it is a useful number to report for
comparison with other studies of spiral galaxies.
We list our measured $S_N$ in Table 7, where we have used the
reddening-corrected $V_{T0}$ galaxy magnitudes from the RC3 catalog
\citep{dev95} and the distances of \citet{karachentsev03} to calculate $M_V$.  We
find $<S_N>\sim$1 for the Sculptor group as a whole, compared to
$<S_N>\sim1.2$ for the late-type galaxies listed in \citet{ashman98} and in \citet{goudfrooij03}.  The Sculptor group galaxies thus appear to have $S_N$ typical for late-type galaxies.  Nevertheless, there is considerable diversity among the four galaxies in our sample, as NGC 253 appears to have the poorest GC system of all comparable late-type galaxies, while NGC 300 has among the richest (Figure \ref{fig13}).

Near-infrared bands, while still affected by the presence of young populations, are known to be more sensitive to the light from the red giants of older populations than are optical bands.  We thus also calculated $K$-band normalized GC specific frequencies, where we defined:
\begin{equation}
S_{N,K} = N_{\rm GC}\times10^{0.4(M_K+18)}
\end{equation}
The zero point of $M_K$ was chosen to account for the typical $V-K=3$ color of E galaxies.  The values
of $S_{N,K}$ are listed in Table 7 and shown in Figure \ref{fig13}, in
comparison with those galaxies in \citet{ashman98} that have $K$-band
magnitudes in \citet{jarrett03}.  For consistency with the way the $S_N$ were derived, we corrected the $K$-band magnitudes using the internal and foreground extinction values from \citet{dev95}, where we took $A_K=0.08A_B$ (Schlegel et al.\ 1998); the largest correction for $K$-band extinction was 0.13 magnitudes.
Normalized to the $K$-band, the
average specific frequency of the \citet{ashman98} and \citet{goudfrooij03} late-type galaxies
is $\sim$50\% higher than if normalized to $V$, while that in elliptical or S0 galaxies is only $\sim$10\% higher, suggesting that the GC
specific frequencies of late-type galaxies are indeed supressed by
contamination from the light of young stars.  Although much less
powerful, the gist of this result is in agreement with
\citet{mclaughlin99}, who found similar GC formation efficiencies in
the Milky Way halo and in giant elliptical galaxies.  The specific
frequency of NGC 253 remains almost unchanged, however, suggesting
that it simply has a very poor GC population.  The specific
frequencies of the other three Sculptor group galaxies are greatly
boosted once normalized in $K$.  Although the error bars are large,
their specific frequencies are more typical of early-type galaxies
than late-type galaxies.

\section{Conclusions}
Previous searches for globular clusters in the Sculptor group have
found 17 GCs in NGC 55 and NGC 253 \citep{dc82,b00}.  Through our
survey, we have discovered additional GCs in NGC 55 and NGC 253 and
confirmed the existence of the first GCs in NGC 247 and NGC 300; in
all, we have found 19 new Sculptor group globular clusters.  We have
also produced Washington $CMT_1$ photometry of 16 of the 17 previously known GCs
and new spectroscopy for 5 of these.  In addition, we have presented
lists of unconfirmed candidate GCs in NGC 45, NGC 55, NGC 247, NGC
253, NGC 300, and NGC 7793; these candidates supplement those
published by \citet{l83a},\citet{l83b}, \citet{b86}, \citet{b00}, \citet{kim02}.

From our photometry, we have found that the bulk of the Sculptor group
GCs are metal-poor.  Our measurements of spectroscopic indices in 6
high signal-to-noise spectra and in the combined spectrum of 18 other
GCs indicate that they have [Fe/H]$\lesssim -1$ and ages consistent
with those of old Milky Way GCs.  The combined luminosity function of
the Sculptor group GCs has high statistical probability of being drawn
from the same parent distribution as the Milky Way GCs.

Despite these strong similarities between the Sculptor group GCs and
those of the Milky Way, our data indicate some important distinctions.
First, our spectroscopic analysis indicates that the [$\alpha$/Fe]
element ratio is $-0.2\pm0.3$ in the Sculptor group GCs, which is 2$\sigma$ lower than
the average [$\alpha$/Fe] ratio in Milky Way GCs.  This may mean that the timescale for GC formation in the Sculptor group was longer than that for the appearance of type Ia supernovae.
Second, all of the
Sculptor group GC systems have kinematics consistent with rotation
with the \ion{H}{1} disk, although only in NGC 253 is this result
based on more than a handful of objects.  \citet{whiting99} presented dynamical evidence that the Sculptor group galaxies acquired their spin in tidal interactions early in their history; the disk kinematics of the GC systems fit with this suggestion, as the GCs must have been present before or formed as a result of these interactions.
Disk kinematics may also be seen
in the old GC system of the LMC (Schommer et al.\ 1992, although note the argument to the contrary by van den Bergh 2004), whose members have ages
within 1 Gyr of those of the oldest Milky Way GCs
\citep{m96,olsen98,johnson99}.  Thus, it may be that the formation
sites of many of the first generation GCs were the disks of galaxies,
rather than halos, as appears to be seen in cosmological simulations
(e.g. Kravtsov et al.\ 2003).  This suggestion would naturally fit
into a picture where halo GCs represent the combination of the GC
systems of many smaller merger fragments, such as in the scenario
proposed by \citet{c98}.  

The kinematics of the  NGC 55 and NGC 300 GC systems are consistent with having circular orbits following the flat rotation curves derived from \ion{H}{1} observations \citep{p90,p91a}.   The kinematics of the NGC 253 GC system show evidence for asymmetric drift; we used this observation to show that the NGC 253 disk at the time of GC system formation had roughly the same scale length as it does now.

The globular cluster specific frequencies ($S_N$) of the Sculptor group galaxies appear similar to those of other late-type galaxies,
although NGC 253 appears unusually poor in GCs, while NGC 300 appears unusually
rich.  We suggest that normalizing the specific frequencies to the near-infrared  luminosities of the host galaxies ($S_{N,K}$), rather than to the traditionally used $V$-band luminosities, is more
appropriate for late-type galaxies.  We find that $S_{N,K}$ of a sample of late-type galaxies is a factor 1.5 larger than their $S_N$, suggesting less contamination by the light from young stars in the infrared.  In NGC 253, we find that $S_{N,K}$ is as low as $S_N$, while in NGC 55, NGC 247, and NGC 300 it is a factor $\gtrsim$2 higher.  Indeed, in NGC 55, NGC 247, and NGC 300, $S_{N,K}$ is more typical of early-type than late-type galaxies.  Similarly large specific frequencies have been found by \citet{seth04} in a sample of Virgo/Fornax dIrr galaxies; thus, some late-type galaxies may be as rich in GCs as are elliptical galaxies.

Finally, Many of our conclusions are tempered by fact that we have discovered only 5$-$20\% of the Sculptor group's globular clusters.  We expect that many more GCs lie projected
against the crowded galaxy disks, which we avoided in our survey; a number may also be
bluer than our inadvertently red color selection.  Other GCs may be in our
candidate lists but not yet confirmed spectroscopically, while a handful may be found at large radii from the host galaxies.  Additional spectroscopy of Sculptor group globular cluster candidates should yield valuable results.

\acknowledgements
We dedicate this paper to the memory of our friend Bob Schommer, who provided much of the inspiration for undertaking this project.  We are greatly saddened by his loss.  We thank Robbie Dohm-Palmer and Mario Mateo for providing their data for use as flat fields, and the CTIO mountain staff for their untiring support of Hydra and Mosaic.  KO thanks Dara Norman for many stimulating discussions.  We very much appreciate the thorough and expedient reading of this paper by the anonymous referee.  This research has made use of the NASA/IPAC Extragalactic Database (NED) which is operated by the Jet Propulsion Laboratory, California Institute of Technology, under contract with the National Aeronautics and Space Administration.

\newpage

\begin{figure}
\plotone{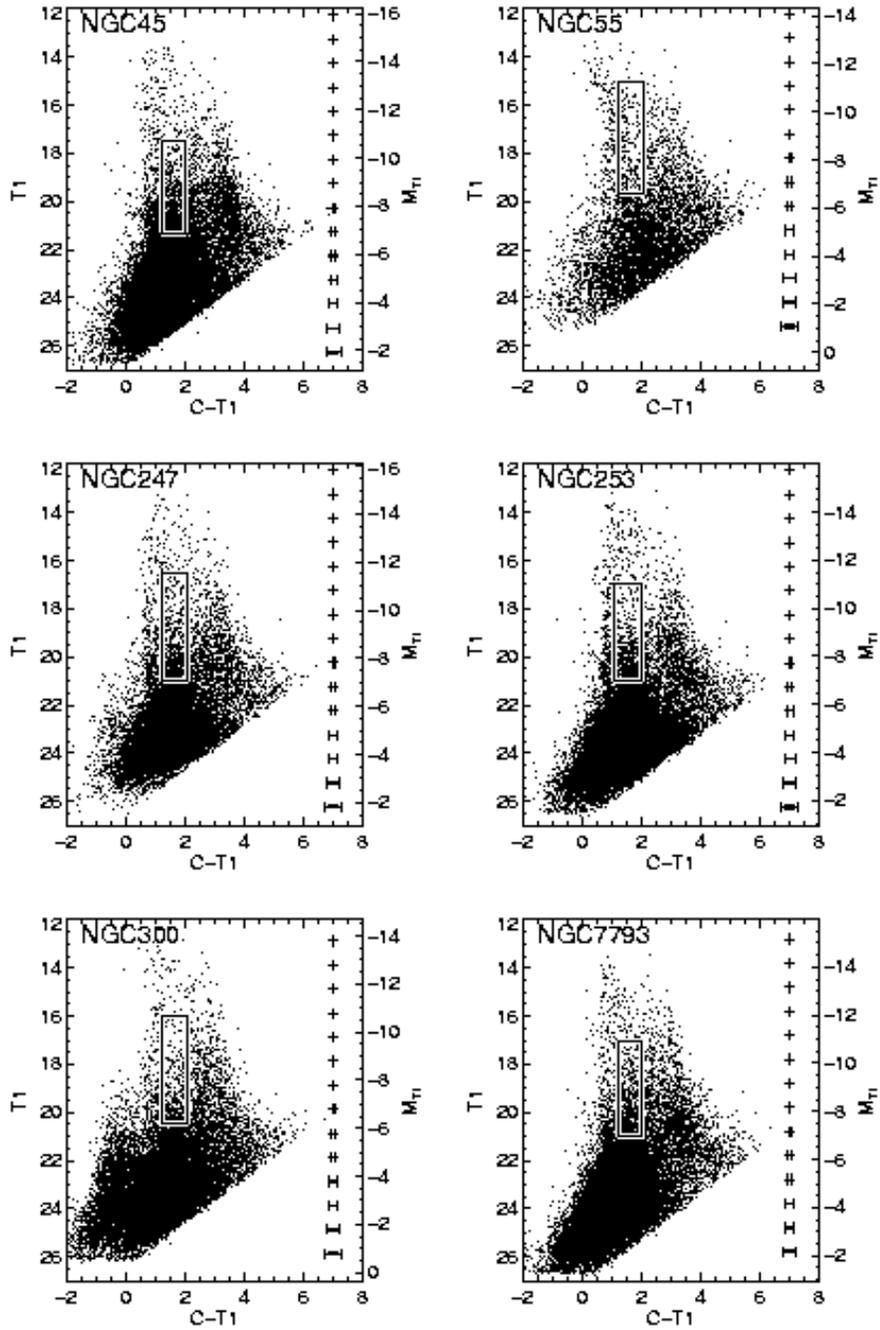}
\caption{$C-T_1,T_1$ color-magnitude diagrams of the image areas outside the ellipses marked in Figure \ref{fig1}, as measured using SExtractor \citep{bertin96}.  The gray boxes mark the regions within which we selected globular cluster candidates.  The right hand axes are labelled with equivalent absolute magnitudes assuming the distances reported by \citet{karachentsev03} for NGC 55, NGC 247, NGC 253, NGC 300, and NGC 7793 and that of \citet{c97} for NGC 45.  The representative error bars contain the SExtractor internal errors only.}
\label{fig2}
\end{figure}

\begin{figure}
\plotone{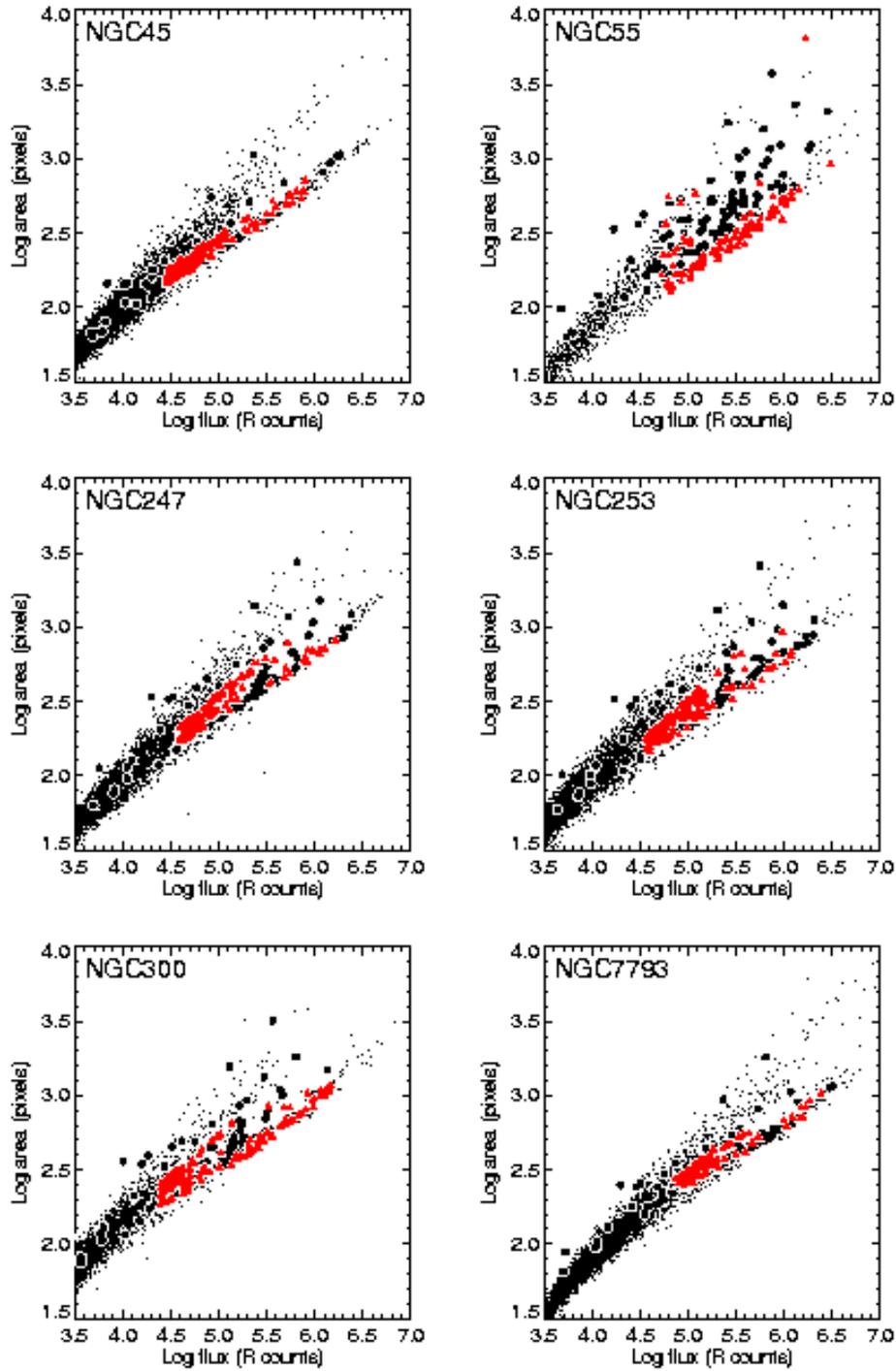}
\caption{These diagrams demonstrate our morphological selection of Sculptor group globular cluster candidates.  The vertical axis shows the logarithm of the number of pixels contained within the isophote having surface brightness 5$\sigma$ above the background; the horizontal axis shows the logarithm of the counts measured through the $R$ filter.  The small points are those objects measured by SExtractor that have ellipticities $e<0.4$ and colors $1.0 \le C-T_1 \le 2.0$.  The filled circles reveal how the Milky Way globular clusters would appear if placed at the distances of the Sculptor group galaxies.  The red triangles show our selected globular cluster candidates.}
\label{fig3}
\end{figure}

\begin{figure}
\epsscale{1.0}
\plotone{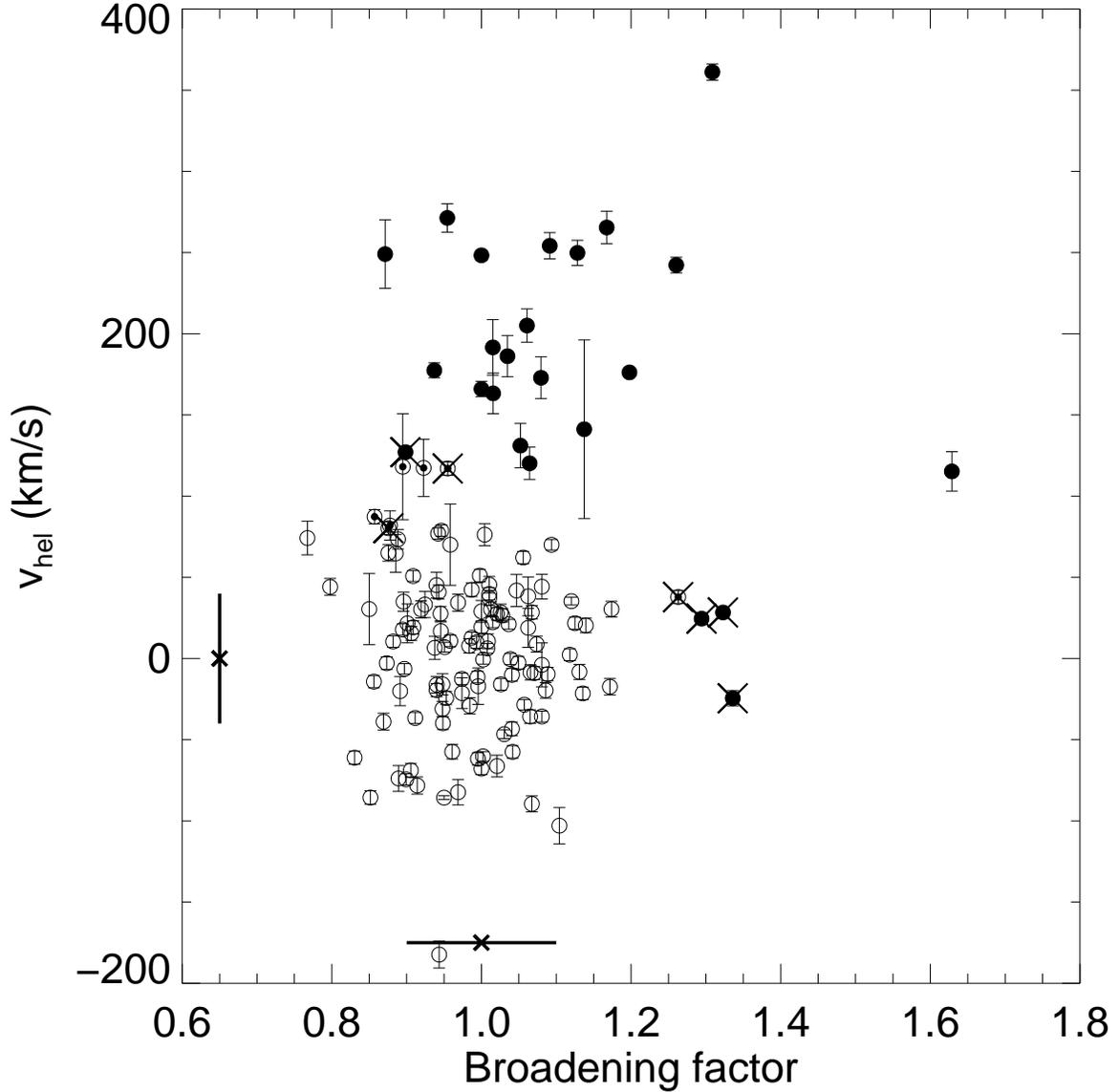}
\caption{Separation of Sculptor group star clusters from foreground stars.  On the horizontal axis, we plot the FWHM of objects in NGC 55, NGC 247, NGC 253, and NGC 300 divided by the mean FWHM of all point-like sources found in the respective images; the cross with horizontal bar at the bottom of the plot shows the mean and 1$\sigma$ standard deviation of the broadening factor. The vertical axis shows our measured heliocentric velocities, while the cross with vertical bar on the left-hand side of the plot shows the mean and 1$\sigma$ standard deviation of the velocity of 568 Galactic foreground stars from \citet{rv00}.  Open circles are objects that cannot be distinguished from foreground stars; open circles with filled central dots are objects that have either velocities or broadening factors in excess of 2$\sigma$ of the stellar means; filled circles have velocities or broadening factors $>$3$\sigma$ times the stellar means.  The points with large crosses indicate objects that we spectroscopically identify as foreground stars.  The remaining 24 broad and/or high-velocity objects we consider to be star clusters in the Sculptor group.}
\label{fig4}
\end{figure}

\begin{figure}
\epsscale{0.8}
\plotone{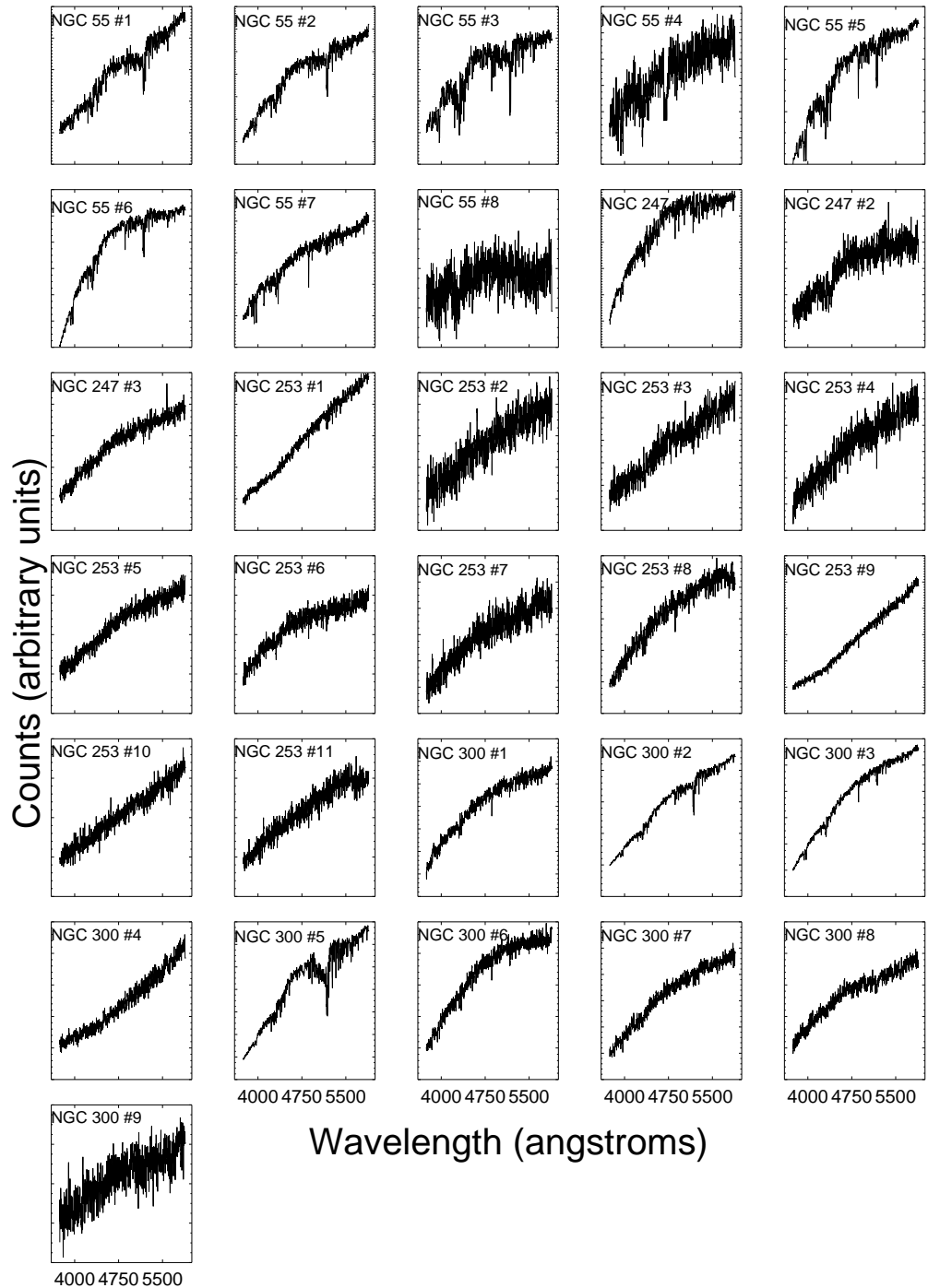}
\caption{Combined spectra of the 31 objects with broadening factors and/or heliocentric velocities in excess of 2$\sigma$ greater than the means for foreground stars.  Seven of the spectra contain the strong MgH, Mg$b$, and Ca 4227 features of foreground dwarfs.  The remaining 24 spectra we consider to belong to Sculptor group star clusters.}
\label{fig5}
\end{figure}

\begin{figure}
\plotone{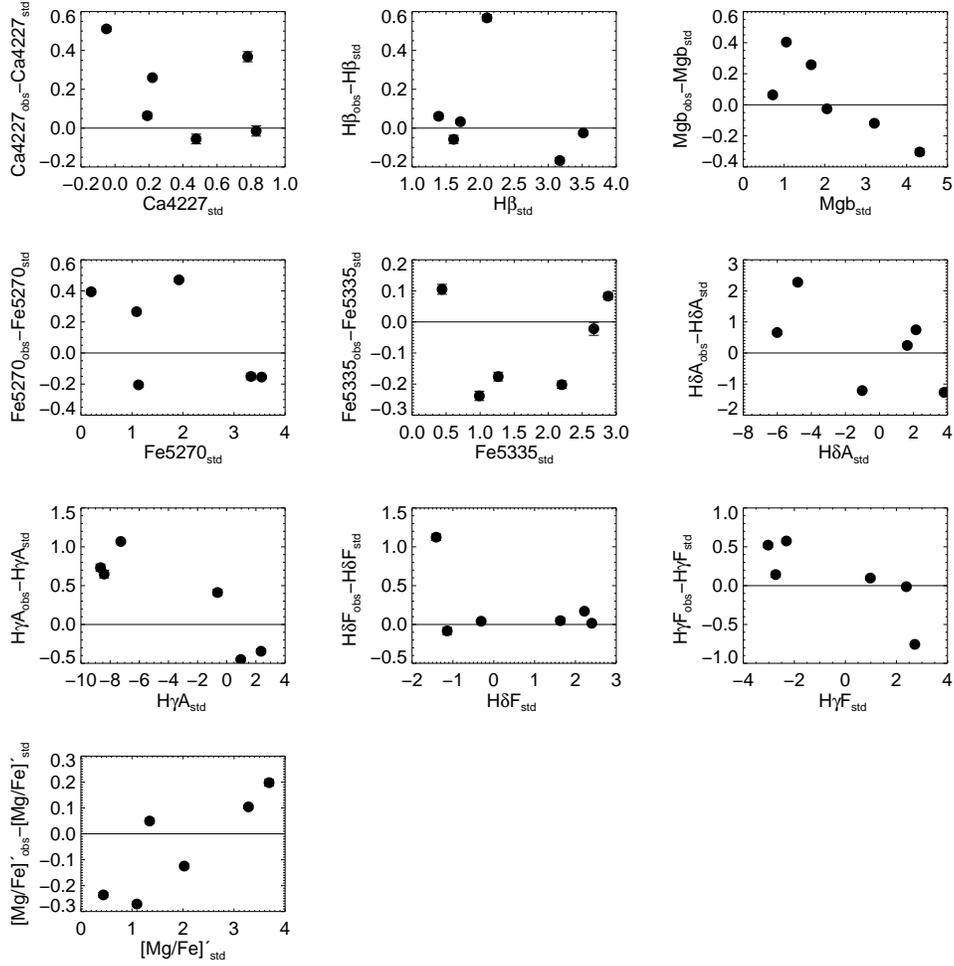}
\caption{Residuals of the measured spectral indices and their standard values for the Lick standard stars HD 3567, HD 22879, HD 23249, HD 196755, HD 218527, HD 219449, and HD 222368.  While there appear to be some systematic trends in the residuals, they are on the whole smaller than the internal scatter; the exception appears to be the Ca4227 index, which has a bias of $\sim$+0.2\AA.}
\label{fig7}
\end{figure}

\begin{figure}
\epsscale{1.0}
\plotone{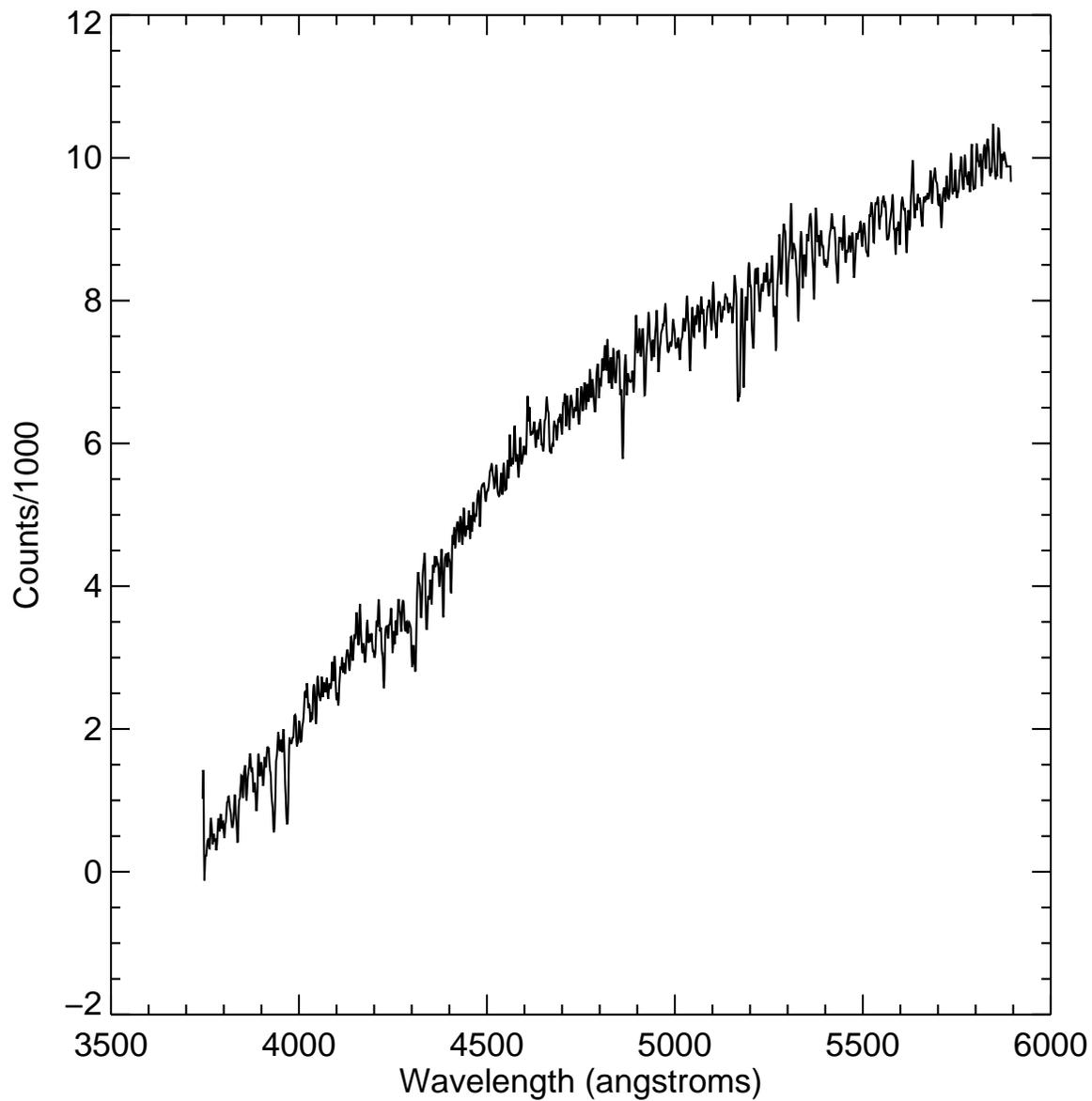}
\caption{The combined spectrum of the 18 Sculptor group star clusters whose individual spectra have signal-to-noise $<$25 per resolution element.}
\label{fig8}
\end{figure}

\begin{figure}
\epsscale{0.7}
\plotone{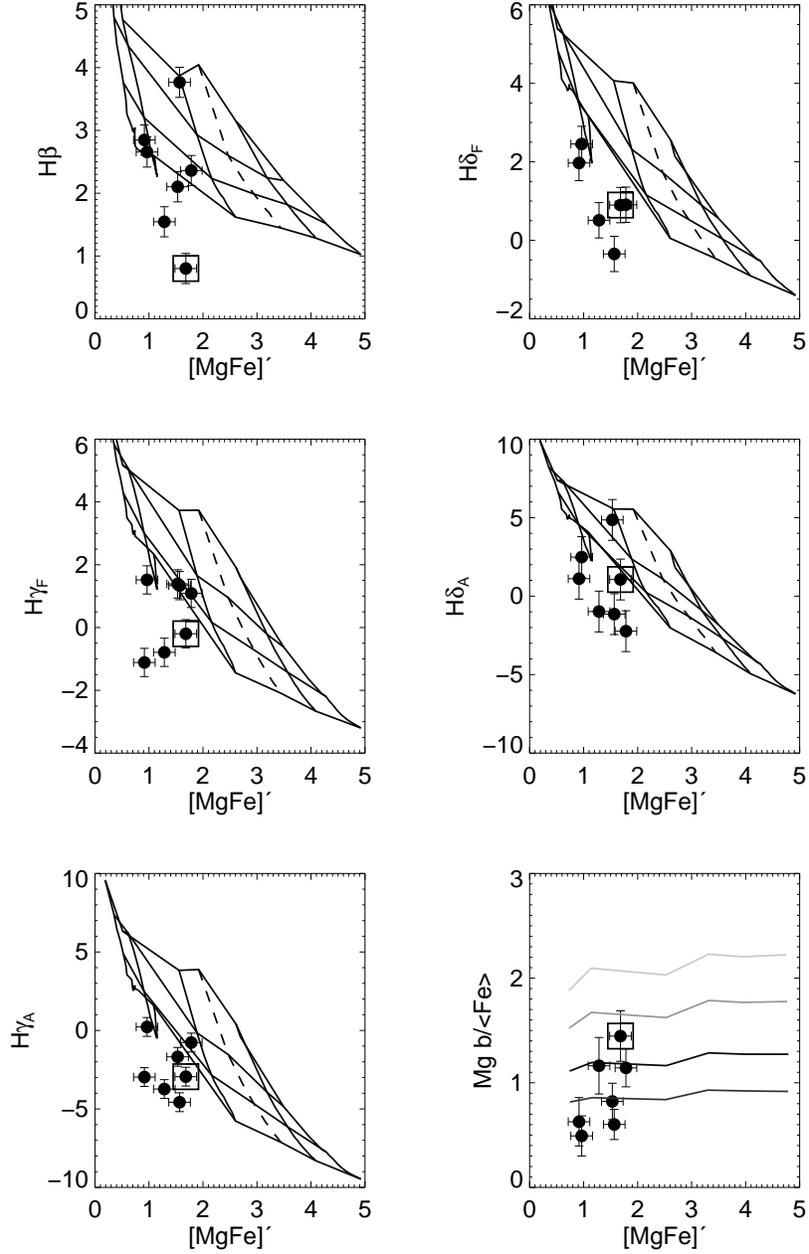}
\caption{The first five panels show age-metallicity diagnostic diagrams for the six Sculptor group star cluster spectra with signal-to-noise $>$25 and for the combined spectrum of 18 lower signal-to-noise spectra (marked by open square).  Overlaid are isochrones and isometallicity tracks from the single stellar population models of Thomas et al.\ (2003).  Metallicity increases from left to right in the diagram, while age increases from top to bottom.  The nearly horizontal lines are isochrones with ages of 1, 2, 5, and 15 Gyr.  The nearly vertical lines are isometallicity tracks with values of [Z/H] of -2.25, -1.35, -0.33, 0.00, +0.35, and +0.67; the dashed line marks the solar metallicity track.  All have [$\alpha$/Fe]=0.0.  The panel at lower right shows the effect of varying [$\alpha$/Fe] on the ratio Mg$b$/$<$Fe$>$ for models with an age of 12 Gyr and metallicities spanning the range $-2.25 \le {\rm [Z/H]} \le +0.67$.  The values of [$\alpha$/Fe] are -0.3 (bottom line), 0.0 (second from bottom line), +0.3 (second from top line), and +0.5 (top line).  Milky Way globular clusters have [$\alpha$/Fe]$\sim$+0.3.}
\label{fig9}
\end{figure}

\begin{figure}
\epsscale{1.0}
\plotone{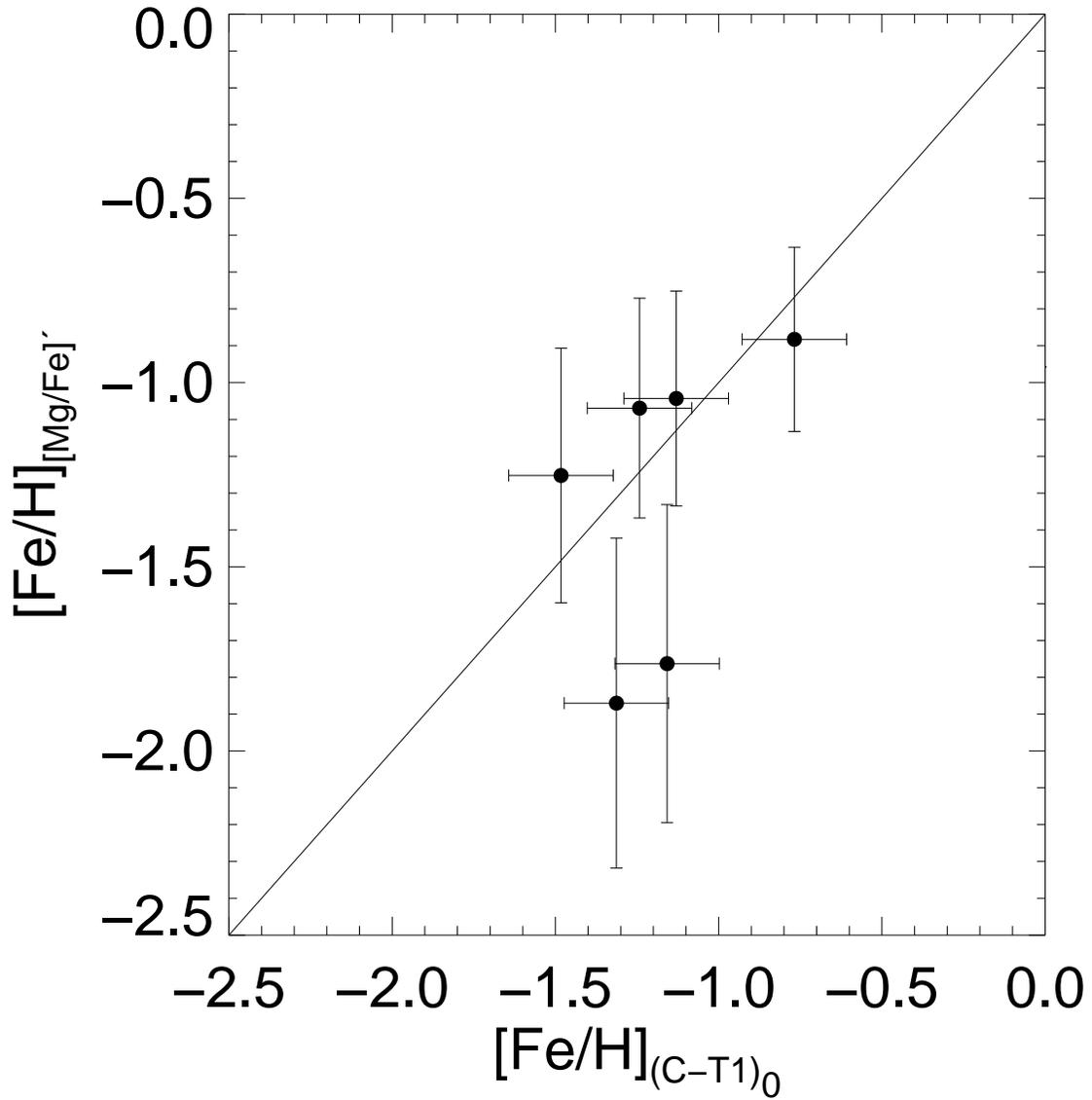}
\caption{Comparison of metallicities derived from the spectroscopic index [Mg/Fe]$^\prime$ and from the $(C-T_1)_\circ$ color of our six Sculptor group globular clusters with high signal-to-noise spectra.  The straight line marks a one-to-one relationship.}
\label{fig10}
\end{figure}

\begin{figure}
\plotone{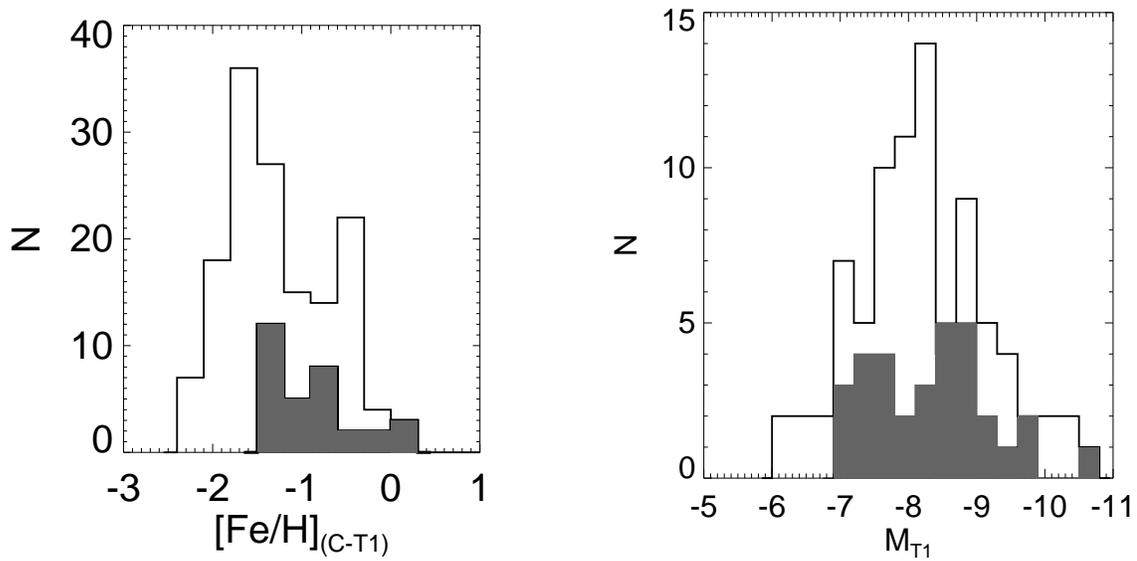}
\caption{Comparison of the metallicity distribution, as derived from our $C-T_1$ photometry, and luminosity function pf the Sculptor group globular clusters (shaded histograms) with the metallicities and luminosities of Milky Way globular clusters \citep[solid lines]{h96}.  }
\label{fig12}
\end{figure}

\begin{figure}
\plotone{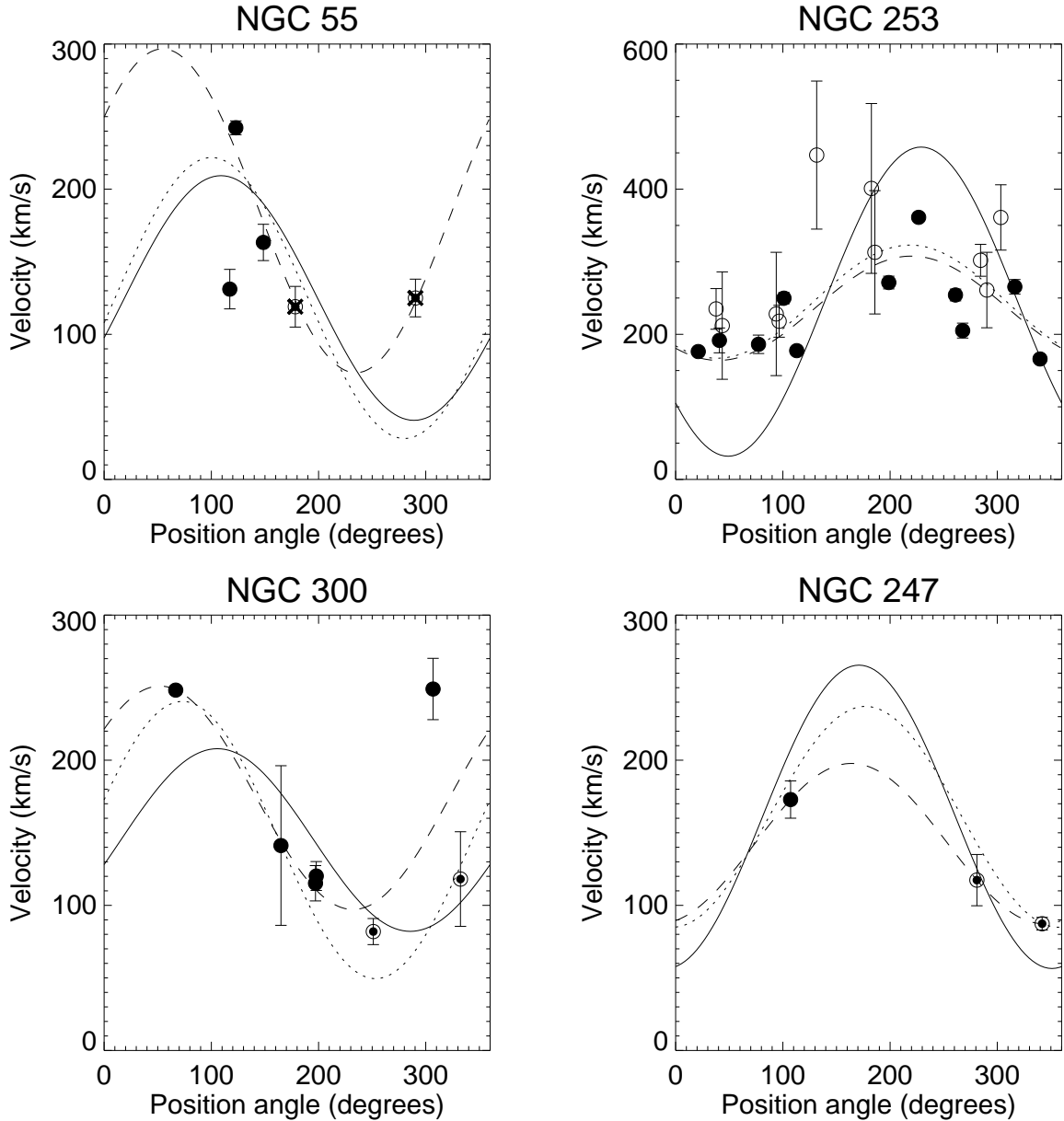}
\caption{Kinematics of the Sculptor group globular cluster systems.  The filled circles (open circles with central dots) represent our measurements for objects with broadening factors and/or heliocentric velocities 3$\sigma$ (2$\sigma$) larger than foreground stars.  Open circles are the globular clusters from \citet{b00}, while crosses are the two NGC 55 globular clusters from \citet{dc82}.  The solid sinusoidal lines are the projected \ion{H}{1} rotation curves of\citep{c90}, \citet{p90}, \citet{p91a}.  The dashed lines are our fits to the globular clusters assuming a flat rotation curve and allowing the systemic velocities, projected rotational velocities, and phase of the projected rotation curve to vary; the dotted lines show the fits in which the systemic velocities were fixed to the values derived from the \ion{H}{1}. 
 }
\label{fig11}
\end{figure}

\begin{figure}
\plotone{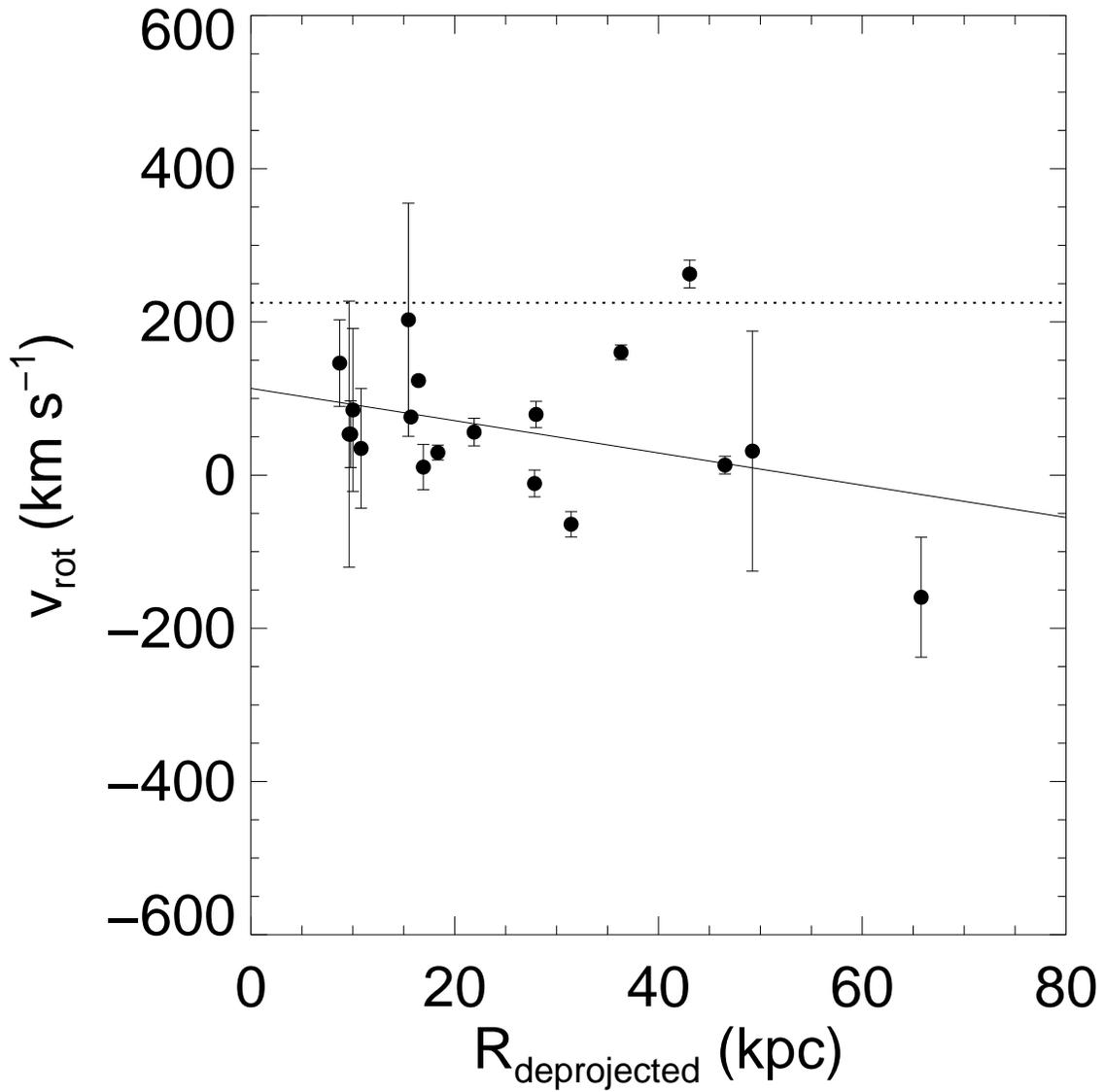}
\caption{Another view of the globular cluster kinematics in NGC 253.  The horizontal axis plots the deprojected galactocentric radii of the globular clusters, assuming that they lie in a disk inclined by 72\arcdeg; the vertical axis shows the deprojected rotational velocities under the same assumptions.  The solid line is a biweighted robust linear fit to the points; the dotted line marks the peak velocity of the \ion{H}{1} rotation curve.}
\label{ngc253vel}
\end{figure}

\begin{figure}
\plotone{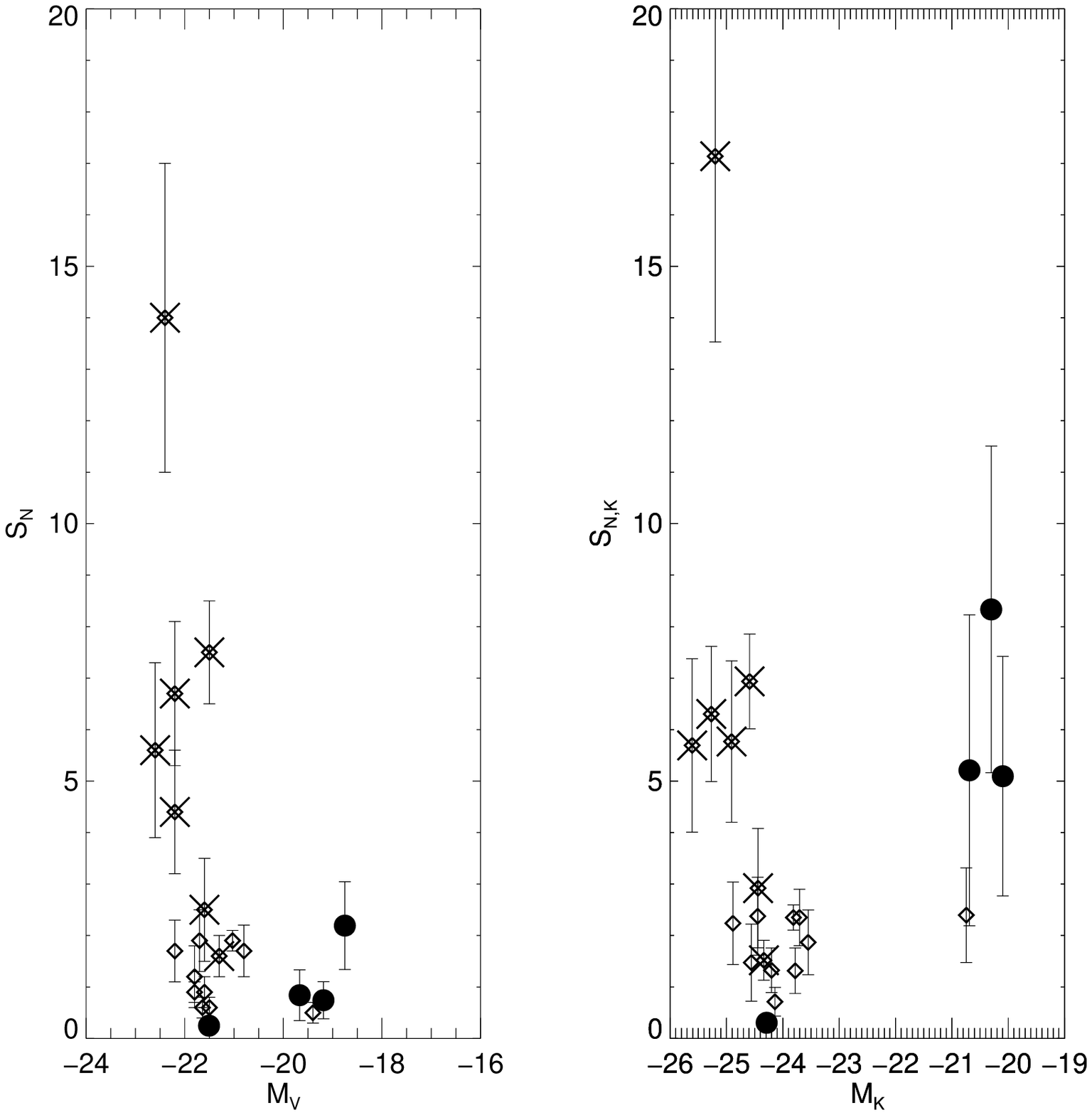}
\caption{Globular cluster specific frequencies in the Sculptor group compared to those in other galaxies.  In the left panel, the optical galaxy magnitudes of \citet{dev95} are used to compute $S_N$.  In the right panel, the 2MASS-based $K$ magnitudes from \citet{jarrett03} are used to calculate $S_{N,K}$ as defined in the text.  The solid circles represent the Sculptor group galaxies, while the crosses are E and S0 galaxies and the open diamonds late-type galaxies from the compilation of \citet{ashman98} and \citet{goudfrooij03}, where we have used only those galaxies with both optical and near-infrared total magnitudes.}
\label{fig13}
\end{figure}

\newpage
\pagestyle{empty}
\begin{deluxetable}{lccrr}
\tablecaption{Log of Observations} 
\tablewidth{0pt}
\tabletypesize{\small}
\tablehead{ 
\colhead{Target} & 
\colhead{Instrument} &
\colhead{Filter} &
\colhead{Date observed} &
\colhead{Exposure time (s)}
}
\startdata
NGC 55 & Mosaic 2 & $C$ & 1999 Nov 11 & 250 \\
NGC 55 & Mosaic 2 & $M$ & 1999 Nov 11 & 250 \\
NGC 55 & Mosaic 2 & $R$ & 1999 Nov 11 & 250 \\
NGC 247 & Mosaic 2 & $C$ & 1999 Nov 11 & 1410 \\
NGC 247 & Mosaic 2 & $M$ & 1999 Nov 11 & 1410 \\
NGC 247 & Mosaic 2 & $R$ & 1999 Nov 11 & 1410 \\
NGC 253 & Mosaic 2 & $C$ & 1999 Nov 11 & 1410 \\
NGC 253 & Mosaic 2 & $M$ & 1999 Nov 11 & 1410 \\
NGC 300 & Mosaic 2 & $C$ & 1999 Nov 11 & 600 \\
NGC 300 & Mosaic 2 & $M$ & 1999 Nov 11 & 600 \\
NGC 300 & Mosaic 2 & $R$ & 1999 Nov 11 & 600 \\
NGC 253 & Mosaic 2 & $R$ & 1999 Nov 12 & 1410 \\
NGC 45 & Mosaic 2 & $C$ & 1999 Nov 12 & 2000 \\
NGC 45 & Mosaic 2 & $M$ & 1999 Nov 12 & 2000 \\
NGC 45 & Mosaic 2 & $R$ & 1999 Nov 12 & 2000 \\
NGC 55 & Mosaic 2 & $C$ & 1999 Nov 12 & 80 \\
NGC 55 & Mosaic 2 & $M$ & 1999 Nov 12 & 80 \\
NGC 55 & Mosaic 2 & $R$ & 1999 Nov 12 & 80 \\
NGC 300 & Mosaic 2 & $C$ & 1999 Nov 12 & 200 \\
NGC 300 & Mosaic 2 & $M$ & 1999 Nov 12 & 200 \\
NGC 300 & Mosaic 2 & $R$ & 1999 Nov 12 & 200 \\
NGC 247 & Mosaic 2 & $C$ & 1999 Nov 12 & 300 \\
NGC 247 & Mosaic 2 & $M$ & 1999 Nov 12 & 300 \\
NGC 247 & Mosaic 2 & $R$ & 1999 Nov 12 & 300 \\
NGC 7793 & Mosaic 2 & $R$ & 1999 Nov 12 & 850 \\
NGC 7793 & Mosaic 2 & $C$ & 1999 Nov 13 & 2550 \\
NGC 7793 & Mosaic 2 & $M$ & 1999 Nov 13 & 2550 \\
NGC 7793 & Mosaic 2 & $R$ & 1999 Nov 13 & 2550 \\
NGC 253 & Mosaic 2 & $C$ & 1999 Nov 13 & 200 \\
NGC 253 & Mosaic 2 & $M$ & 1999 Nov 13 & 200 \\
NGC 253 & Mosaic 2 & $R$ & 1999 Nov 13 & 200 \\
HD 3567 & Hydra-CTIO/Air Schmidt & \nodata & 2000 Oct 31 & 300 \\
HD 22879 & Hydra-CTIO/Air Schmidt & \nodata &  2000 Oct 31 & 300 \\
HD 23249 & Hydra-CTIO/Air Schmidt & \nodata &  2000 Oct 31 & 25 \\
HD 196755 & Hydra-CTIO/Air Schmidt & \nodata &  2000 Nov 2 & 110 \\
HD 218527 & Hydra-CTIO/Air Schmidt & \nodata &  2000 Nov 2 & 80 \\
HD 219449 & Hydra-CTIO/Air Schmidt & \nodata &  2000 Nov 2 & 50 \\
HD 222368 & Hydra-CTIO/Air Schmidt & \nodata &  2000 Nov 2 & 50 \\
NGC 55 & Hydra-CTIO/400mm & \nodata & 2001 Oct 17 & 13500 \\
HD195633 & Hydra-CTIO/400mm & \nodata & 2001 Oct 17 & 10 \\
HD223311 & Hydra-CTIO/400mm & \nodata & 2001 Oct 17 & 5 \\
NGC 253 & Hydra-CTIO/400mm & \nodata & 2001 Oct 18 & 16200 \\
HD223311 & Hydra-CTIO/400mm & \nodata & 2001 Oct 18 & 5 \\
NGC 300 & Hydra-CTIO/400mm & \nodata & 2001 Oct 19 & 16200 \\
HD217877 & Hydra-CTIO/400mm & \nodata & 2001 Oct 19 & 3 \\
HD223311 & Hydra-CTIO/400mm & \nodata & 2001 Oct 19 & 90 \\
NGC 247 & Hydra-CTIO/400mm & \nodata & 2001 Oct 20 & 13500 \\
47 Tuc & Hydra-CTIO/400mm & \nodata & 2001 Oct 20 & 600 \\
HD223311 & Hydra-CTIO/400mm & \nodata & 2001 Oct 20 & 10 \\
\enddata 
\end{deluxetable}

\begin{deluxetable}{lccc}
\tablecaption{Measured photometric transformation coefficients}
\tablewidth{0pt}
\tablehead{\colhead{Chip \#} &
\colhead{$A_0$} & \colhead{$A_1$} & \colhead{$A_2$}
}
\startdata
1 &  0.505(14) & -0.142(11) & 0.32(3) \\
2 &  0.497(20) & -0.141(15) & 0.32(3) \\
3 &  0.494(22) & -0.153(17) & 0.32(3) \\
4 &  0.492(17) & -0.149(9) & 0.32(3) \\
5 &  0.467(24) & -0.159(29) & 0.32(3) \\
6 &  0.473(9) & -0.147(11) & 0.32(3) \\
7 &  0.516(15) & -0.163(20) & 0.32(3) \\
8 &  0.489(13) & -0.140(18) & 0.32(3) \\
\tableline
&  $B_0$ &  $B_1$ &  $B_2$  \\
\tableline
1 & -0.884(16) & -0.160(17) & 0.19(2) \\
2 & -0.853(11) & -0.141(17) & 0.19(2) \\
3 & -0.856(14) & -0.128(12) & 0.19(2) \\
4 & -0.856(15) & -0.122(24) & 0.19(2) \\
5 & -0.867(15) & -0.133(23) & 0.19(2) \\
6 & -0.852(13) & -0.150(13) & 0.19(2) \\
7 & -0.863(16) & -0.123(21) & 0.19(2) \\
8 & -0.864(12) & -0.119(16) & 0.19(2) \\
\tableline
&  $C_0$ &  $C_1$ &  $C_2$  \\
\tableline
1 & -0.649(7) & -0.018(5) & 0.14(3) \\
2 & -0.640(13) & -0.020(10) & 0.14(3) \\
3 & -0.633(6) & -0.016(4) & 0.14(3) \\
4 & -0.623(15) & -0.010(12) & 0.14(3) \\
5 & -0.657(11) & -0.016(9) & 0.14(3) \\
6 & -0.608(15) & -0.026(3) & 0.14(3) \\
7 & -0.623(6) & -0.024(4) & 0.14(3) \\
8 & -0.644(8) & -0.012(6) & 0.14(3) \\
\enddata
\end{deluxetable}

\begin{deluxetable}{lccccccccccccr}
\rotate
\tablecaption{Candidate Sculptor group star clusters}
\tablewidth{0pt}
\tabletypesize{\small}
\tablehead{\colhead{Number} &
\colhead{R.A.} & \colhead{Dec.} & \colhead{log area} & \colhead{$e$} &
\colhead{$C$} & \colhead{$\sigma_C$} & 
\colhead{$M$} & \colhead{$\sigma_M$} & 
\colhead{$T_1$} & \colhead{$\sigma_{T_1}$} &
\colhead{$v_{\rm hel}$} & \colhead{$\sigma_{v_{\rm hel}}$} & \colhead{Note} \\
& (2000.0) & & (pixel$^2$) & & & & & & & & (km s$^{-1}$) & (km s$^{-1}$) \\
}
\startdata
NGC 247 \\
\tableline
   1 &          0:45:48.756 &         -20:27:12.19 &   2.32 &   0.08 &     21.927 &      0.017 &     21.260 &      0.009 &     20.550 &      0.007 &    \nodata &    \nodata &     no spectrum \\
   2 &          0:45:51.535 &         -20:46:55.88 &   2.28 &   0.13 &     22.035 &      0.019 &     21.291 &      0.009 &     20.606 &      0.007 &    \nodata &    \nodata &   weak spectrum \\
   3 &          0:45:51.652 &         -20:59:51.26 &   2.37 &   0.18 &     22.445 &      0.025 &     21.508 &      0.011 &     20.563 &      0.007 &    \nodata &    \nodata &     no spectrum \\
   4 &          0:45:52.098 &         -21:03:36.51 &   2.45 &   0.08 &     20.828 &      0.009 &     20.159 &      0.005 &     19.549 &      0.004 &    \nodata &    \nodata &     no spectrum \\
   5 &          0:45:52.528 &         -20:36:06.72 &   2.35 &   0.13 &     22.366 &      0.022 &     21.407 &      0.010 &     20.481 &      0.007 &    \nodata &    \nodata &     no spectrum \\
   6 &          0:45:52.971 &         -20:53:13.42 &   2.32 &   0.10 &     22.778 &      0.029 &     21.752 &      0.013 &     20.781 &      0.008 &    \nodata &    \nodata &     no spectrum \\
   7 &          0:45:54.488 &         -20:31:44.65 &   2.33 &   0.05 &     22.077 &      0.020 &     21.406 &      0.010 &     20.416 &      0.006 &    \nodata &    \nodata &     no spectrum \\
   8 &          0:45:54.994 &         -20:41:01.00 &   2.29 &   0.13 &     22.410 &      0.024 &     21.805 &      0.014 &     20.741 &      0.008 &    \nodata &    \nodata &     no spectrum \\
   9 &          0:45:55.280 &         -20:33:19.49 &   2.43 &   0.04 &     21.196 &      0.011 &     20.605 &      0.006 &     19.936 &      0.005 &    \nodata &    \nodata &     no spectrum \\
  10 &          0:45:57.405 &         -20:56:32.82 &   2.40 &   0.08 &     21.692 &      0.015 &     21.035 &      0.008 &     20.365 &      0.006 &    \nodata &    \nodata &          galaxy \\
......\\
\enddata
\end{deluxetable}

\begin{deluxetable}{llccccccccc}
\rotate
\tablecaption{Properties of Sculptor group star clusters}
\tablewidth{0pt}
\tabletypesize{\small}
\tablehead{\colhead{Name\tablenotemark{a}} & \colhead{Original\tablenotemark{b}} & \colhead{Number\tablenotemark{c}} &
\colhead{$M_{T_1}$} & \colhead{$E(B-V)$} & \colhead{$(C-T_1)_0$} & \colhead{[Fe/H]$_{(C-T_1)_0}$} &
\colhead{[Fe/H]$_{\rm Lick}$} & 
\colhead{$v_{\rm hel}$} & \colhead{Broad} & \colhead{Significance} \\
\colhead{} & \colhead{Name} & \colhead{} &\colhead{(mag)} & \colhead{(mag)} & \colhead{(mag)} & \colhead{} &
\colhead{} & 
\colhead{(km s$^{-1}$)} & \colhead{} & \colhead{($\sigma$)}
}
\startdata
   NGC55-4 & &    83 &     -7.4 &    0.015 &    1.249 &    -1.38 $\pm$     0.16 &              \nodata &        131.1 $\pm$     13.6 &   1.05 &   3 \\
   NGC55-7 & LA43&    27 &     -9.2 &    0.014 &    1.341 &    -1.16 $\pm$     0.16 &     -1.76 $\pm$ 0.43 &    242.3 $\pm$      4.8 &   1.26 &   3 \\
   NGC55-8 & &    71 &     -7.0 &    0.015 &    1.468 &    -0.88 $\pm$     0.16 &              \nodata &        163.3 $\pm$     12.5 &   1.02 &   3 \\
  NGC247-1 & &    64 &     -9.7 &    0.018 &    1.353 &    -1.13 $\pm$     0.16 &     -1.04 $\pm$ 0.29 &        87.3  $\pm$      4.4 &   0.86 &   2 \\
  NGC247-2 & &    53 &     -7.5 &    0.018 &    1.725 &    -0.41 $\pm$     0.16 &              \nodata &        117.4 $\pm$     17.6 &   0.92 &   2 \\
  NGC247-3 & &    73 &     -8.5 &    0.018 &    1.460 &    -0.89 $\pm$     0.16 &              \nodata &        172.9 $\pm$     12.9 &   1.08 &   3 \\
  NGC253-1 & LA40&    46 &     -9.6 &    0.019 &    1.522 &    -0.77 $\pm$     0.16 &     -0.88 $\pm$ 0.25 &    166.0 $\pm$      4.7 &   1.00 &   3 \\
  NGC253-2 & &    26 &     -8.5 &    0.019 &    1.330 &    -1.18 $\pm$     0.16 &              \nodata &        361.2 $\pm$      5.0 &   1.31 &   3 \\
  NGC253-3 & &    92 &     -8.2 &    0.019 &    1.585 &    -0.65 $\pm$     0.16 &              \nodata &        249.8 $\pm$      7.7 &   1.13 &   3 \\
  NGC253-4 & &   114 &     -7.6 &    0.020 &    1.571 &    -0.67 $\pm$     0.16 &              \nodata &        191.6 $\pm$     17.1 &   1.02 &   3 \\
  NGC253-5 & B15&    19 &     -8.4 &    0.019 &    1.271 &    -1.33 $\pm$     0.16 &              \nodata &     205.1 $\pm$     10.3 &   1.06 &   3 \\
  NGC253-6 & &   109 &     -8.0 &    0.019 &    1.286 &    -1.29 $\pm$     0.16 &              \nodata &        177.5 $\pm$      4.6 &   0.94 &   3 \\
  NGC253-7 & &    73 &     -7.6 &    0.019 &    1.845 &    -0.23 $\pm$     0.16 &              \nodata &        176.2 $\pm$      2.4 &   1.20 &   3 \\
  NGC253-8 & &   110 &     -8.6 &    0.019 &    1.264 &    -1.34 $\pm$     0.16 &              \nodata &        186.2 $\pm$     12.7 &   1.03 &   3 \\
  NGC253-9 & LA35&     5 &    -10.6 &    0.019 &    1.305 &    -1.24 $\pm$     0.16 &     -1.07 $\pm$ 0.30 &    254.2 $\pm$      8.1 &   1.09 &   3 \\
 NGC253-10 & LA57&    15 &     -8.6 &    0.020 &    1.665 &    -0.51 $\pm$     0.16 &              \nodata &    265.4 $\pm$     10.0 &   1.17 &   3 \\
 NGC253-11 & &    48 &     -8.6 &    0.019 &    1.530 &    -0.75 $\pm$     0.16 &              \nodata &        271.3 $\pm$      8.8 &   0.95 &   3 \\
  NGC300-1 & &    48 &     -8.1 &    0.013 &    1.276 &    -1.31 $\pm$     0.16 &     -1.87 $\pm$ 0.45 &        115.2 $\pm$     12.2 &   1.63 &   3 \\
  NGC300-3 & &   121 &     -8.9 &    0.011 &    1.209 &    -1.48 $\pm$     0.16 &     -1.25 $\pm$ 0.35 &        248.3 $\pm$      3.1 &   1.00 &   3 \\
  NGC300-4 & &    23 &     -7.3 &    0.015 &    1.936 &    -0.12 $\pm$     0.16 &              \nodata &        249.1 $\pm$     21.1 &   0.87 &   3 \\
  NGC300-6 & &    29 &     -7.2 &    0.013 &    1.208 &    -1.48 $\pm$     0.16 &              \nodata &        81.9  $\pm$      9.0 &   0.88 &   2 \\
  NGC300-7 & &    44 &     -7.6 &    0.016 &    1.284 &    -1.29 $\pm$     0.16 &              \nodata &        120.2 $\pm$     10.0 &   1.06 &   3 \\
  NGC300-8 & &    63 &     -7.6 &    0.013 &    2.069 &     0.02 $\pm$     0.16 &              \nodata &        141.2 $\pm$     55.0 &   1.14 &   3 \\
  NGC300-9 & &    32 &     -7.0 &    0.014 &    1.486 &    -0.84 $\pm$     0.16 &              \nodata &        118.1 $\pm$     32.6 &   0.89 &   2 \\
  DG2      & &       &     -8.8 &    0.013 &    1.379(40) & -1.07 $\pm$     0.16 &              \nodata &    119$\pm$14\tablenotemark{d}\\
  DG3      & &       &     -8.9 &    0.013 &    1.246(40) & -1.39 $\pm$     0.16 &              \nodata &    125$\pm$13\tablenotemark{d}\\
  BS7      & B1&     &      -7.7 &    0.019 &    0.370(52) & \nodata &              \nodata &                212$\pm$   74\tablenotemark{e} \\
  BS8      & LA26&   &	   -8.7	&    0.019 &	1.273(22) & -1.3202$\pm$     0.16 &              \nodata &   235$\pm$   28\tablenotemark{e}\\
  BS28 	   & LA11&   &	   -8.4	&    0.019 &	0.739(14) & \nodata &              \nodata & 		     228$\pm$   85\tablenotemark{e}\\
  BS40 	   & LA24&   &	   -9.7	&    0.019 &	1.571(80) & -0.6741$\pm$     0.16 &              \nodata &   401$\pm$  117\tablenotemark{e}\\
  BS42 	   & B24&    &	   -8.4	&    0.019 &	1.248(54) & -1.3834$\pm$     0.16 &              \nodata &   313$\pm$   85\tablenotemark{e}\\
  BS44 	   & LA3&    &	   -9.3	&    0.019 &	2.052(28) &  0.0029$\pm$     0.16 &              \nodata &   447$\pm$  102\tablenotemark{e}\\
  BS57 	   & B12&    &	   -8.8	&    0.019 &	1.402(31) & -1.0181$\pm$     0.16 &              \nodata &   361$\pm$   45\tablenotemark{e}\\
  BS58 	   & B13&    &	   -7.5	&    0.019 &	2.972(319)&  \nodata &              \nodata & 		     261$\pm$   52\tablenotemark{e}\\
  BS61 	   & B14&    &	   -8.9	&    0.019 &	0.717(32) & \nodata         &     \nodata & 		     302$\pm$   22\tablenotemark{e}\\
\enddata	     
\tablenotetext{a}{DG$\equiv$\citet{dc82}; BS$\equiv$\citet{b00}}
\tablenotetext{b}{B$\equiv$\citet{b86}; LA$\equiv$\citet{l83a,l83b}}
\tablenotetext{c}{Corresponds to number in column 1 of Table 3}
\tablenotetext{d}{Velocity from \citet{dc82}}
\tablenotetext{e}{Velocity from \citet{b00}}
\end{deluxetable}

\begin{deluxetable}{lcccccccccc}
\tablecaption{Lick indices}
\tablewidth{0pt}
\tabletypesize{\small}
\tablehead{\colhead{ID} &
\colhead{Ca4227} & \colhead{H$\beta$} & \colhead{H$\delta_F$} & \colhead{H$\gamma_F$} & \colhead{H$\delta_A$} & \colhead{H$\gamma_A$} & \colhead{Mgb} & \colhead{Fe5270} & \colhead{Fe5335} & \colhead{[MgFe]$^\prime$} \\
\colhead{} &
\colhead{(\AA)} & \colhead{(\AA)} & \colhead{(\AA)} & \colhead{(\AA)} & \colhead{(\AA)} & \colhead{(\AA)} & \colhead{(\AA)} & \colhead{(\AA)} & \colhead{(\AA)} & \colhead{(\AA)}}
\startdata
   NGC55-7 &   0.056 &   2.656 &   2.454 &   1.518 &   2.487 &   0.223 &   0.679 &   1.353 &   1.406 &   0.964 \\ 
$\pm$ &   0.230 &   0.240 &   0.450 &   0.450 &   1.300 &   0.600 &   0.250 &   0.310 &   0.150 &   0.200 \\ 
  NGC247-1 &   0.147 &   3.766 &  -0.348 &   1.335 &  -1.146 &  -4.572 &   1.130 &   2.546 &   1.211 &   1.567 \\ 
 &   0.230 &   0.240 &   0.450 &   0.450 &   1.300 &   0.600 &   0.250 &   0.310 &   0.150 &   0.200 \\ 
  NGC253-1 &   1.730 &   2.361 &   0.906 &   1.088 &  -2.232 &  -0.771 &   2.007 &   1.365 &   2.149 &   1.783 \\ 
 &   0.230 &   0.240 &   0.450 &   0.450 &   1.300 &   0.600 &   0.250 &   0.310 &   0.150 &   0.200 \\ 
  NGC253-9 &   1.179 &   2.102 &   6.695 &   1.390 &   4.856 &  -1.681 &   1.357 &   1.825 &   1.478 &   1.531 \\ 
 &   0.230 &   0.240 &   0.450 &   0.450 &   1.300 &   0.600 &   0.250 &   0.310 &   0.150 &   0.200 \\ 
  NGC300-1 &   0.404 &   2.848 &   1.970 &  -1.113 &   1.110 &  -2.970 &   0.738 &   1.079 &   1.274 &   0.915 \\ 
 &   0.230 &   0.240 &   0.450 &   0.450 &   1.300 &   0.600 &   0.250 &   0.310 &   0.150 &   0.200 \\ 
  NGC300-3 &   0.727 &   1.545 &   0.510 &  -0.790 &  -0.982 &  -3.734 &   1.379 &   1.210 &   1.162 &   1.285 \\ 
 &   0.230 &   0.240 &   0.450 &   0.450 &   1.300 &   0.600 &   0.250 &   0.310 &   0.150 &   0.200 \\ 
   Low S/N &   0.802 &   0.801 &   0.897 &  -0.198 &   1.061 &  -2.945 &   2.114 &   1.181 &   1.743 &   1.682 \\ 
 ~spectra (combined) &   0.230 &   0.240 &   0.450 &   0.450 &   1.300 &   0.600 &   0.250 &   0.310 &   0.150 &   0.200 \\ 

\enddata
\end{deluxetable}

\begin{deluxetable}{lrrrrrrrrrr}
\tablecaption{Sculptor group globular cluster kinematics}
\tablewidth{0pt}
\tablehead{\colhead{Galaxy} & \colhead{$\sigma_{v_{\rm los}}$} & \colhead{$v_{\rm sys}$} & \colhead{$v_{\rm rot,proj}$} & \colhead{$\theta_{\rm rot}$}  & \colhead{$\theta_{\rm rot}-\theta_{\rm HI}$} & \colhead{$(v/\sigma)_{\rm rot}$} \\
\colhead{} & \colhead{(km s$^{-1}$)} & \colhead{(km s$^{-1}$)} & \colhead{(km s$^{-1}$)} & \colhead{(\arcdeg)}  & \colhead{(\arcdeg)} & \colhead{}}
\startdata
    NGC 55 &       45.7 &      185.1 $\pm$ 22.6 &      111.6 $\pm$ 45.6&      -35.3 $\pm$ 25.3 &      -54.3 $\pm$ 25.3 &        2.4 $\pm$ 1.0 \\
 & &      $\equiv$125.0 &       96.8 $\pm$ 39.1 &       10.2 $\pm$ 25.7&       -8.8 $\pm$ 25.7 &        1.4 $\pm$ 0.6 \\
   NGC 247 &       35.4 &      142.5 $\pm$ 10.0 &       55.2 $\pm$ 10.9&       73.9 $\pm$ 16.8 &       -7.1 $\pm$ 16.8 &  \nodata \\
 & &      $\equiv$161.0 &       76.1 $\pm$ 15.6 &       87.1 $\pm$ 12.0&        6.1 $\pm$ 12.0 &        6.3 $\pm$  1.3\\
   NGC 253 &       76.1 &      235.9 $\pm$  8.8 &       71.8 $\pm$ 21.5&      129.8 $\pm$ 17.9 &       -9.2 $\pm$ 17.9 &        1.2 $\pm$ 0.4 \\
 & &      $\equiv$245.0 &       77.9 $\pm$ 18.1 &      128.6 $\pm$ 13.6&      -10.4 $\pm$ 13.6 &        1.3 $\pm$ 0.3 \\
   NGC 300 &       62.4 &      174.1 $\pm$ 11.6 &       77.0 $\pm$ 19.6&      -37.9 $\pm$ 15.0 &      -53.5 $\pm$ 15.0 &        2.3 $\pm$ 0.6 \\
 & &      $\equiv$145.0 &       95.5 $\pm$ 21.4 &      -16.9 $\pm$ 13.3&      -32.5 $\pm$ 13.3 &        2.5 $\pm$ 0.6 \\
 \enddata
\end{deluxetable}

\begin{deluxetable}{lccccccccc}
\tablecaption{Sculptor group globular cluster specific frequencies}
\tablewidth{0pt}
\tablehead{\colhead{Galaxy} & \colhead{$N_{\rm GC}$} & \colhead{Completeness} & \colhead{$S_N$} & \colhead{$S_{N,K}$}
}
\startdata
    NGC 55 &     5 &   0.14 &   0.74 $\pm$   0.36 &   5.09 $\pm$   2.33 \\ 
   NGC 247 &     3 &   0.05 &   0.84 $\pm$   0.49 &   5.21 $\pm$   3.02 \\ 
   NGC 253 &    21 &   0.21 &   0.25 $\pm$   0.06 &   0.30 $\pm$   0.07 \\ 
   NGC 300 &     7 &   0.10 &   2.19 $\pm$   0.85 &   8.33 $\pm$   3.17 \\ 
\enddata
\end{deluxetable}

\end{document}